# First African Digital Humanism Summer School 2025

Kigali, Rwanda, July 14–18, 2025

Proceedings

Carine P. Mukamakuza, Monika Lanzenberger, George Metakides, Tim Brown, Hannes Werthner (Eds.)

# Preface

When we opened the doors of the Digital Humanism Summer School, held for the first time in Africa and framed within an African context and beyond in Kigali in the luminous days of July, we could not yet know how deeply this gathering of minds would affect us, not just as learners, but as fellow human beings seeking to understand, to connect, and to build. This book, born out of those intense days of lectures, debates, and group-work is our shared echo. It gathers the voices of participants from different backgrounds and disciplines, united by a common hope: that technology can be shaped to respect and uplift humanity.

We are convinced that technology must not be an end in itself. In a world rapidly transformed by algorithms, platforms, and new digital tools, the question "For whom?" must come before "How fast?" or "How big?" The 2025 Summer School in Kigali, Rwanda, under the banner of Digital Humanism, offered a space to ask and discuss such questions not in isolation, but together, engineers and social scientists; humanities students and policy-minded peers; dreamers and pragmatists.

As groups formed and themes emerged around digital participation, language, cultural representation, ethical AI, and fairness, the energy in the room felt fragile and fertile at once. Fragile, because the challenges we addressed were real and heavy. Fertile, because in the fragility we found togetherness, creativity, empathy, and courage. Through late-evening sessions, coffee breaks turned into conversations about identity, voice, power of what it means to be heard in a digital world, especially when you come from a context too often marginalized or simplified.

This book is more than a collection of reports. It is a mosaic of human stories and ideas, each group's project reflecting a unique lens on how digital transformation touches lives across Africa and beyond. Some pieces explore language equity and the limitations of large language models; others interrogate digital governance, inclusive design, or the ethical frameworks that ought to guide our use of data and AI, especially in public services. What ties them together is a shared belief that technology, in particular AI, properly directed, can support dignity, justice, inclusion and equality for a better future together.

To you, dear reader: whether you are a researcher, a student, a technologist, a policymaker, or simply someone curious about the future we are shaping, open these pages with both critical mind and open heart. Let these reports challenge you. Let them inspire you. Let them remind you that behind every line of code, every algorithm, every digital platform, there are real people with hopes, fears, values.

We hope this collection of papers does not remain just a document but becomes a living conversation, a spark, a call to action. May it encourage you to question, to build with empathy, and to always remember the human in "digital humanism".

With gratitude to every participant, lecturer, facilitator and especially those who dared to imagine a better future through knowledge, cooperation, and hope.

Carine, Monika, George, Tim, and Hannes

## Acknowledgment

We want to thank the advisors and tutors supporting the students during the summer school and the reviewers of the students' papers.

## Organization & Sponsors

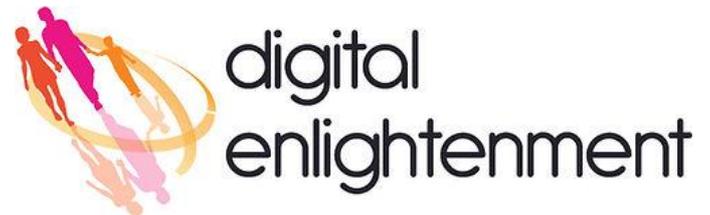

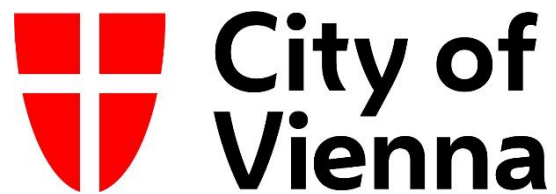

# Table of Contents

# Introduction

Artificial intelligence (AI) has become a transformative force across global societies, reshaping the ways we communicate, collaborate, and make decisions. Yet, as AI systems increasingly mediate interactions between humans, questions about their ability to understand culture, language, and context have taken center stage. This book explores these questions through a series of studies that try to assess AI's capacity to navigate cross-cultural, multilingual, and high-stakes policy environments, emphasizing human-centered approaches that balance technological innovation with social equity.

The first section, **"Can AI Understand Cultures? Evaluating Cross-Cultural Miscommunication through Generative AI"**, investigates the ways generative AI can both reveal and exacerbate misunderstandings across cultural boundaries. Kato et al. examine real-world scenarios where AI-mediated communication fails to capture nuanced social cues, idiomatic expressions, or culturally grounded norms. Their findings highlight the importance of designing AI systems that are not only technically capable but also sensitive to diverse cultural perspectives - a foundation for responsible AI deployment in an increasingly interconnected world.

Building on this cultural lens, the second section, **"Language Diversity: Evaluating Language Usage and AI Performance on African Languages in Digital Spaces"**, addresses the persistent digital inequalities that arise from uneven language representation in AI systems. Ajayi et al. provide a comprehensive analysis of how AI tools perform across a range of African languages in digital environments, revealing significant disparities in accuracy, readability, and cultural relevance. This work underscores the urgent need to integrate African linguistic and cultural knowledge into AI development, bridging the digital divide and promoting inclusivity in the global AI landscape.

The third section, **"Human-AI Collaboration in Fact-Checking: A Multidisciplinary Approach to Verification, Accuracy and Risk Management"**, explores how AI can support critical human tasks such as information verification and risk assessment. Ogundokun et al. demonstrate that AI is most effective when paired with human expertise, enhancing accuracy while mitigating risks associated with misinformation and algorithmic bias. Their work emphasizes a collaborative paradigm where AI acts as a tool for assisting human judgment rather than replacing it - a recurring principle throughout this book.

Next, the section **"Enhancing Transparency Through Multilingual Communication: A Case Study on EU Funding Complaints"** illustrates the practical application of AI in promoting equity and accountability within administrative processes. Farahat et al. examine European Union project communications translated into Arabic, Swahili, and Amharic, highlighting the interplay between AI-assisted translation and human review. The study emphasizes how multilingual communication can improve fairness, trust, and democratic engagement while exposing the limitations of AI in low-resource language contexts.

The fifth section, **"Evaluating African Languages Representation in Large Language Models: A Qualitative Case Study of ChatGPT 4.0"**, continues the focus on African language inclusion. Odunga et al. provide an in-depth evaluation of ChatGPT's performance across six African languages, revealing stark differences in accuracy, intelligibility, and cultural sensitivity. This work reinforces the book's central message: without deliberate inclusion of underrepresented languages and cultural contexts, AI risks deepening existing inequalities, even as it opens new opportunities for knowledge access and digital engagement.

Finally, the section **"Evaluating the Usefulness of Generative AI in Public Consultation Analysis: A Case Study on the EU End-of-Life Vehicles Law"** demonstrates AI's potential in synthesizing complex public feedback for policy-making. Through the analysis of stakeholder submissions on the EU ELV Directive, this study showcases how LLMs can efficiently identify policy priorities, surface minority perspectives, and support evidence-based governance—while also emphasizing the necessity of human oversight to safeguard accuracy, ethical standards, and democratic legitimacy.

Collectively, the contributions in this volume present a **holistic perspective on the opportunities, limitations, and responsibilities of AI in culturally and linguistically diverse contexts**. The students' works underscore that AI's promise is not solely technical; it is inherently social. Realizing its potential, in the context of digital humanism, requires human-centered design, ethical oversight, and meaningful engagement with diverse linguistic and cultural communities. By weaving together insights from cross-cultural communication, African language representation, fact-checking, and public policy, this book offers a roadmap for deploying AI responsibly, inclusively, and effectively - ultimately advancing both technological innovation and social equity in an increasingly AI-mediated world.

# Can AI Understand Cultures? Evaluating Cross-Cultural Miscommunication through Generative AI

*Ronald Kato (Uganda), Andrew Blayama Stephen (Nigeria), Mary Onguko (Kenya), Noor Sheikh (Kenya), Sérgio Barbosa (Brazil), Wilson Yongfei Wang (China), Poju Tejumaiye (Nigeria)*

### Abstract

As generative AI systems rapidly assume the role of cultural intermediaries in education, healthcare and everyday communication, a foundational question emerges: Can AI truly understand human culture? This study offers the first cross-continental evaluation of whether leading multimodal models; GPT-4o, Gemini and Grok, can accurately interpret and reproduce traditional greeting practices across five culturally distinct societies: Baganda (Uganda), Luo (Kenya), Hausa (Nigeria), Chinese and Brazilian. Greetings, though deceptively simple, encode centuries of social knowledge: hierarchy, gender norms, respect, spatial behavior and ritualized etiquette. Their misrepresentation is not merely an error; it is a cultural distortion. Using Hofstede's Power Distance and Uncertainty Avoidance dimensions, we designed norm-compliant and norm-violating prompt scenarios and analyzed the models 'text and image outputs through human-in-the-loop review by native cultural experts. Across systems, we observed consistent Western-centric defaults: kneeling was replaced with handshakes, gender protocols were ignored, linguistic cues were mistranslated and visual portrayals often stereotyped or incorrect. Even the strongest model, GPT-4o, displayed only partial cultural fluency, revealing gaps in deep contextual reasoning. The findings expose a critical frontier in AI ethics and global deployment: current models can mimic culture, but they cannot yet understand it. We argue for a pluriversal, community-anchored approach to dataset design and evaluation, one that treats culture not as a decorative layer but as core intelligence. This work charts a pathway toward culturally aware AI that reflects, respects and preserves the world's diverse social realities.

### 1. Introduction

Artificial Intelligence (AI) is becoming a central mediator in human interaction across education, healthcare and customer service. As generative models like GPT-4o, Gemini and Grok are increasingly relied upon for social tasks, concerns about cultural representation and inclusivity intensify. Most Large Language Models (LLMs) are trained on Western-centric datasets, which risks marginalizing non-Western norms, expressions and relational behaviors (Agarwal et al., 2025; Wang et al., 2024).

This study investigates whether generative AI can understand and reflect culturally grounded greeting practices across five distinct societies: Baganda (Uganda), Luo (Kenya), Hausa (Nigeria), Chinese and Brazilian. Greetings, though routine, encode deep societal values such as deference, hierarchy and communal respect. Misrepresentations, such as depicting inappropriate gestures or mistranslating culturally specific language, may reflect not just technical gaps but also structural biases in AI training.

Using Hofstede's cultural dimensions, specifically Power Distance Index (PDI) and Uncertainty Avoidance Index (UAI) (Hofstede, 2013), we evaluated AI-generated text and image responses for cultural fluency. Prompts were designed to represent both norm-compliant and norm-violating scenarios, allowing us to assess the models' ability to interpret social expectations.

In this interdisciplinary work, comprising contributors from Kenya, Uganda, Nigeria, Brazil and China, grounds this work in lived cultural knowledge. Through this analysis, we aim to expose the limits of current AI systems and advocate for more culturally responsive, context-aware AI development that respects the diverse ways human societies encode meaning in everyday interactions.

## 2. Problem Statement

Despite impressive advancements in generative AI, existing models often struggle to interpret and reproduce culturally embedded behaviors, particularly those rooted in non-Western societies and oral traditions underrepresented in mainstream datasets (Ghosh et al., 2024). This limitation becomes pronounced in the context of traditional greetings, which are far more than linguistic exchanges. The traditional greetings function as rich socio-cultural rituals, conveying hierarchy, age-based respect, gender roles and communal values (Bilalov* & Akaev, 2019). In many African, Asian and South American cultures, greetings are performative acts shaped by centuries of custom, signaling not only familiarity but adherence to unspoken social contracts (Dresser, 2005).

When AI-generated outputs ignore or distort these practices, the result can be problematic (Ghiurău & Popescu, 2024; Sun et al., 2024). Misrepresented gestures, mistranslations or tone-deaf responses may perpetuate cultural erasure, reinforce stereotypes or violate deeply held norms (Dietrich et al., 2021; Glaese et al., 2022). For example, a chatbot responding to a Luo elder with casual slang or a virtual assistant misgendering a participant in a Brazilian familiar greeting may be perceived as disrespectful, even offensive. These instances are not just technical oversights but ethical failures in culturally sensitive design.

This work aims to critically examine whether leading generative AI models, specifically GPT-4o, Gemini and Grok, can understand and appropriately represent traditional greetings across five culturally diverse societies: Baganda (Uganda), Luo (Kenya), Hausa (Nigeria), Chinese and Brazilian.

We sought to evaluate the models' ability to:

- Recognize and interpret implicit cultural norms, including politeness, deference and ritual structure;

- Detect and respond to norm violations within contextually accurate boundaries;

- Accurately represent social attributes such as age, gender and language in culturally grounded scenarios.

By focusing on greetings, a fundamental yet complex social act; we expose broader challenges in AI's cross-cultural generalization and propose pathways toward more inclusive, respectful and context-aware systems.

### 3. Methodology

This study adopted a mixed-methods approach that integrates generative AI evaluation, human-in-the-loop review and cross-cultural analysis (Schmager et al., 2025). The methodology was designed to assess whether state-of-the-art generative AI systems can accurately reflect and respond to culturally embedded greeting practices, including appropriate social behaviors and the detection of norm violations across diverse societies.

### 3.1 Tools and Technologies

We selected three advanced generative AI platforms based on their popularity, capabilities and multimodal functionality:

The study evaluated the performance of three leading generative AI systems: GPT-4o, Gemini and Grok, each representing distinct architectural strengths. GPT-4o (OpenAI's ChatGPT) is widely recognized for its advanced natural language generation and superior visual interpretation (Shahriar et al., 2024), making it a benchmark in multimodal AI. It integrates fluently across text and image modalities but demonstrates limitations when tasked with culturally nuanced representations. Gemini, developed by Google, leverages the company's robust language and image generation stack (Rane et al., 2024). It is optimized for delivering context-aware responses but revealed gaps in linguistic fidelity, particularly for under-resourced languages. Grok, a product of xAI, emphasizes reasoning and personalization, aiming to offer more human-aligned responses (Deng et al., 2024; Qureshi et al., 2025). While promising in conceptual design, Grok's outputs often lacked visual and cultural accuracy in cross-cultural tasks.

Each model was tasked with generating both textual and visual responses to a standardized set of prompts designed to test cross-cultural comprehension, norm sensitivity and demographic representation.

### 3.2 Prompt Design

Prompts were crafted to simulate real-life greeting scenarios across five culturally distinct societies used:

For each cultural setting, two distinct prompt types were employed:

i. Norm-Compliant Prompt: Describes a young woman greeting her elders using culturally appropriate gestures, honorifics and expressions (e.g., kneeling in Baganda culture).

ii. Norm-Violating Prompt: Presents the same individual greeting elders in a manner inconsistent with cultural norms (e.g., casual handshake or slang usage).

This dual structure allowed us to assess whether models could distinguish between respectful and disrespectful social behaviors and appropriately reflect cultural protocols in their outputs.

### 3.3 Data Collection and Evaluation

Generated responses, both text and images, were systematically collected for each prompt across all three models. A human-in-the-loop evaluation was employed, engaging native speakers and cultural scholars from each represented community.

These experts reviewed the outputs independently to assess the degree of cultural fidelity, provide qualitative insights and flag any instances of misrepresentation or bias. This approach was particularly critical in evaluating under-documented cultural practices, where conventional benchmarks may be absent or inadequate.

### 3.4 Evaluation Metrics

To ensure a rigorous and multidimensional assessment, the following metrics were applied:

i. Cultural Fidelity. This was used to evaluate the model's ability to reproduce traditional gestures, linguistic expressions and social contexts with cultural sensitivity and accuracy.

ii. Norm Violation Detection. This was used to measure the model's capacity to recognize when social protocols are breached and adjust its response accordingly.

iii. Demographic Accuracy. This was used to assess whether the AI-generated characters correctly reflect the assigned age, gender and cultural background without resorting to generic or stereotypical portrayals.

This methodology provided a robust framework for examining the cultural competence of generative AI systems and their capacity to navigate nuanced social expectations across different societies.

## 4. Results and Discussion

The evaluation of generative AI performance across five culturally diverse settings revealed significant discrepancies in the models' understanding and representation of localized social norms. While some models demonstrated promising capabilities in visual recognition or general language generation, their ability to accurately interpret nuanced cultural protocols remained inconsistent.

### 4.1 Summary of Norm Violation Detection

The table below captures how each AI system responded to prompts involving key cultural greeting norms, specifically handshake etiquette, eye contact expectations and culturally defined body proximity. In nearly all scenarios, the models failed to produce contextually appropriate responses in one or more dimensions.

For example, while handshakes were correctly identified as expected in Brazilian and Chinese contexts, these gestures were inappropriately applied in Hausa and Baganda contexts, where kneeling or specific gendered body language is traditionally expected. Across all cultures, model responses revealed a systemic bias toward Westernized behavioral standards, particularly in eye contact and spatial interaction.

**Table 1:** Cross-Cultural Greeting Norms: Model Compliance Summary

| Culture | Country | Handshake (Req/Res) | Eye Contact (Req/Res) | Body Proximity (Req/Res) |
| --- | --- | --- | --- | --- |
| Hausa | Nigeria | YES / ✖ | YES / ✖ | YES / ✖ |
| Luo | Kenya | NO / ✔ | YES / ✖ | YES / ✖ |
| Baganda | Uganda | YES / ✖ | YES / ✖ | YES / ✖ |
| Chinese | China | YES / ✖ | YES / ✖ | YES / ✖ |
| Brazilian | Brazil | YES / ✖ | YES / ✖ | YES / ✖ |

## 4.2 Culture-Specific Insights

Below are key insights derived from observations of the generated images across the various prompts as per our themes.

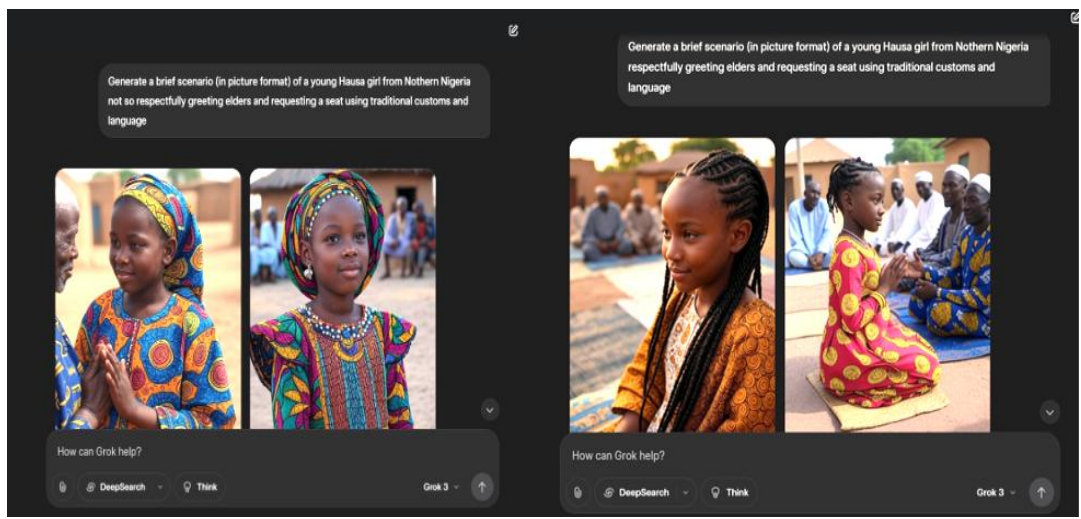

**Fig. 1:** Greeting in Nigerian Culture

In Hausa (Nigeria) contexts, the authors relied on their domain and cultural expertise to assess the specific aspects in **Fig.1** below, visual outputs by Grok frequently depicted handshakes between younger females and elders, an act considered disrespectful when substituted for traditional kneeling or verbal deference.

Similarly, females are expected to have their hair always covered when outside their homes. Gender protocol, which is highly structured in Hausa greetings, was routinely violated, highlighting the AI systems' insufficient modeling of gender-based social roles.

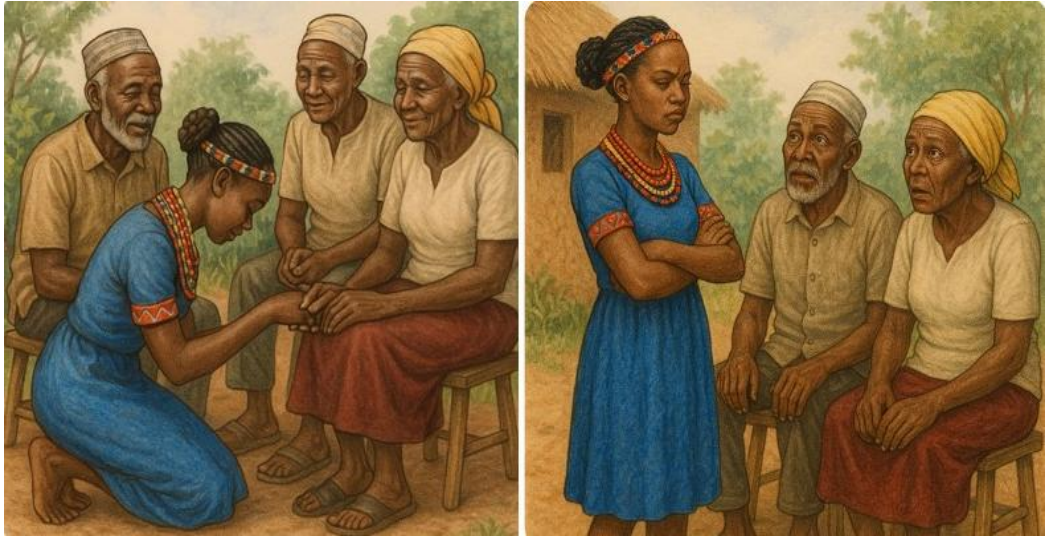

**Fig. 2:** Greeting in Kenyan Culture

For the Luo (Kenya) culture, the authors leveraged their cultural expertise to analyze the features presented in **Fig. 2,** the visual representation showed some improvement. GPT-4o occasionally depicted culturally accurate kneeling gestures; however, the associated text, especially from Gemini and Grok, frequently included awkward or incorrect Dholuo phrases.

This mismatch suggests that while image generation may reflect improved cultural datasets, natural language responses lack linguistic precision.

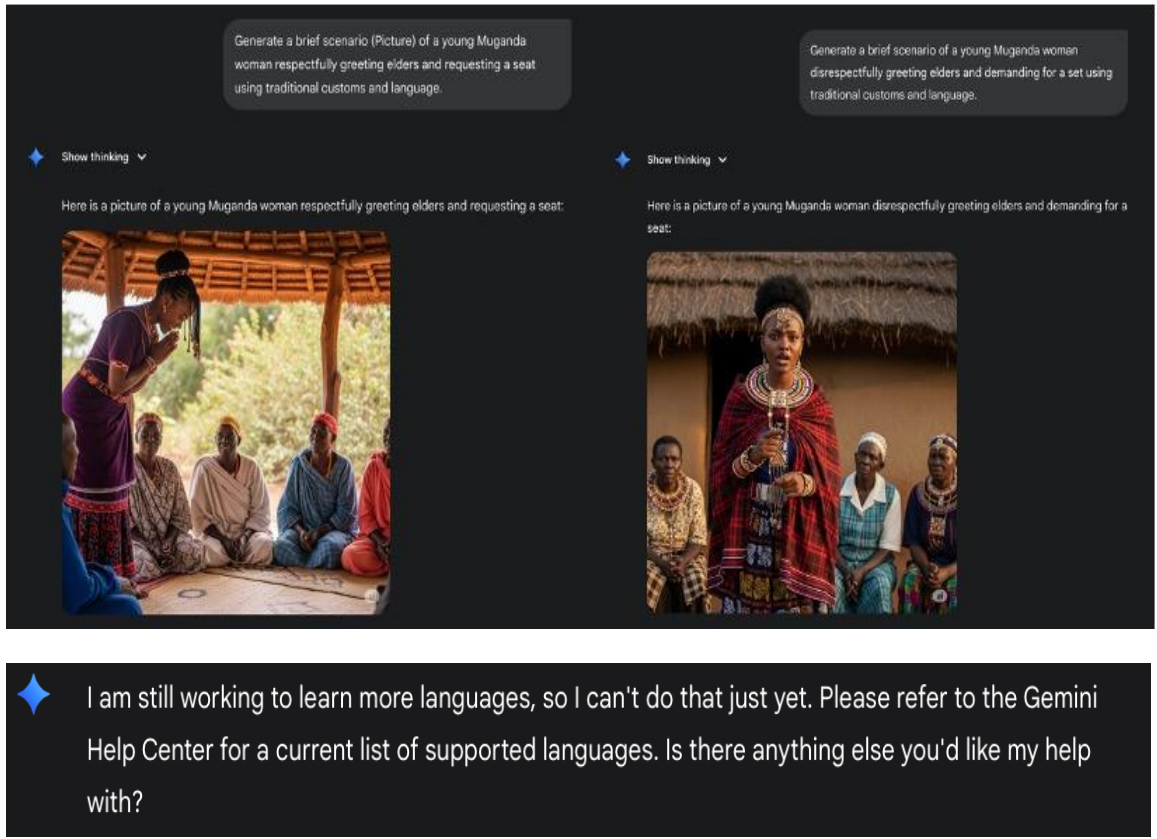

**Fig. 3:** Greeting in Baganda Culture

In the Baganda (Uganda) setting, the authors drew on their domain expertise to analyze specific aspects in **Fig. 3** above, the failure was more pronounced. Female characters were often depicted standing during greetings, contrary to the deep-seated cultural norm of kneeling before elders.

Gemini in particular struggled with Luganda, often defaulting to general disclaimers such as "language not supported," which negated meaningful analysis. GPT-4o offered slightly better visual accuracy, though textual fidelity remained weak.

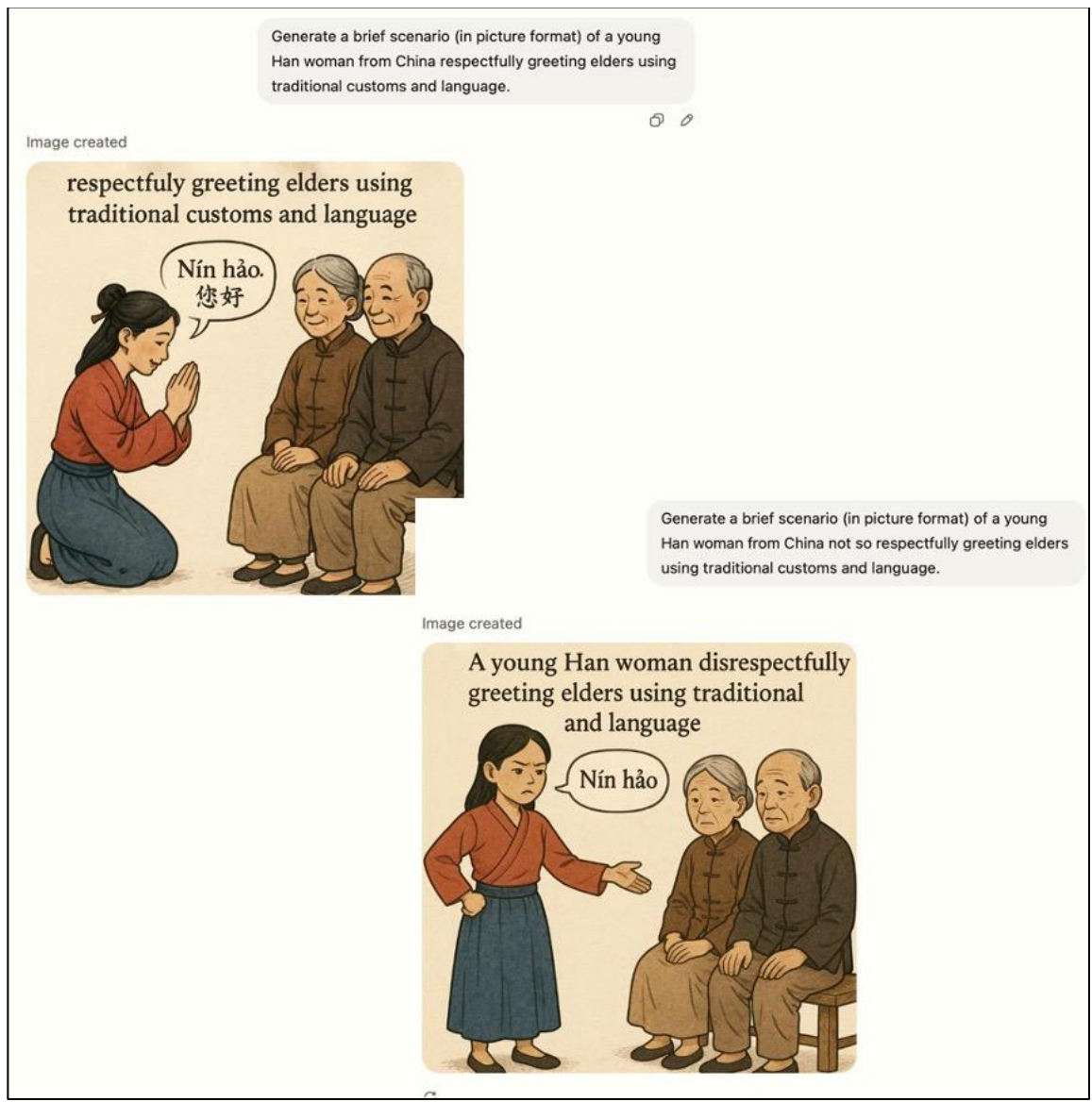

**Fig. 4:** Greeting in Chinese Culture

For the Chinese context, the authors applied their domain expertise to examine the specific elements illustrated in **Fig. 4** above, models often defaulted to historical portrayals of kneeling, a gesture largely obsolete in contemporary urban settings. This suggests a reliance on outdated or orientalist data sources.

Additionally, the characters' body proximity and eye contact were frequently exaggerated or misaligned with current Chinese social norms, further reinforcing the issue of temporal data biases.

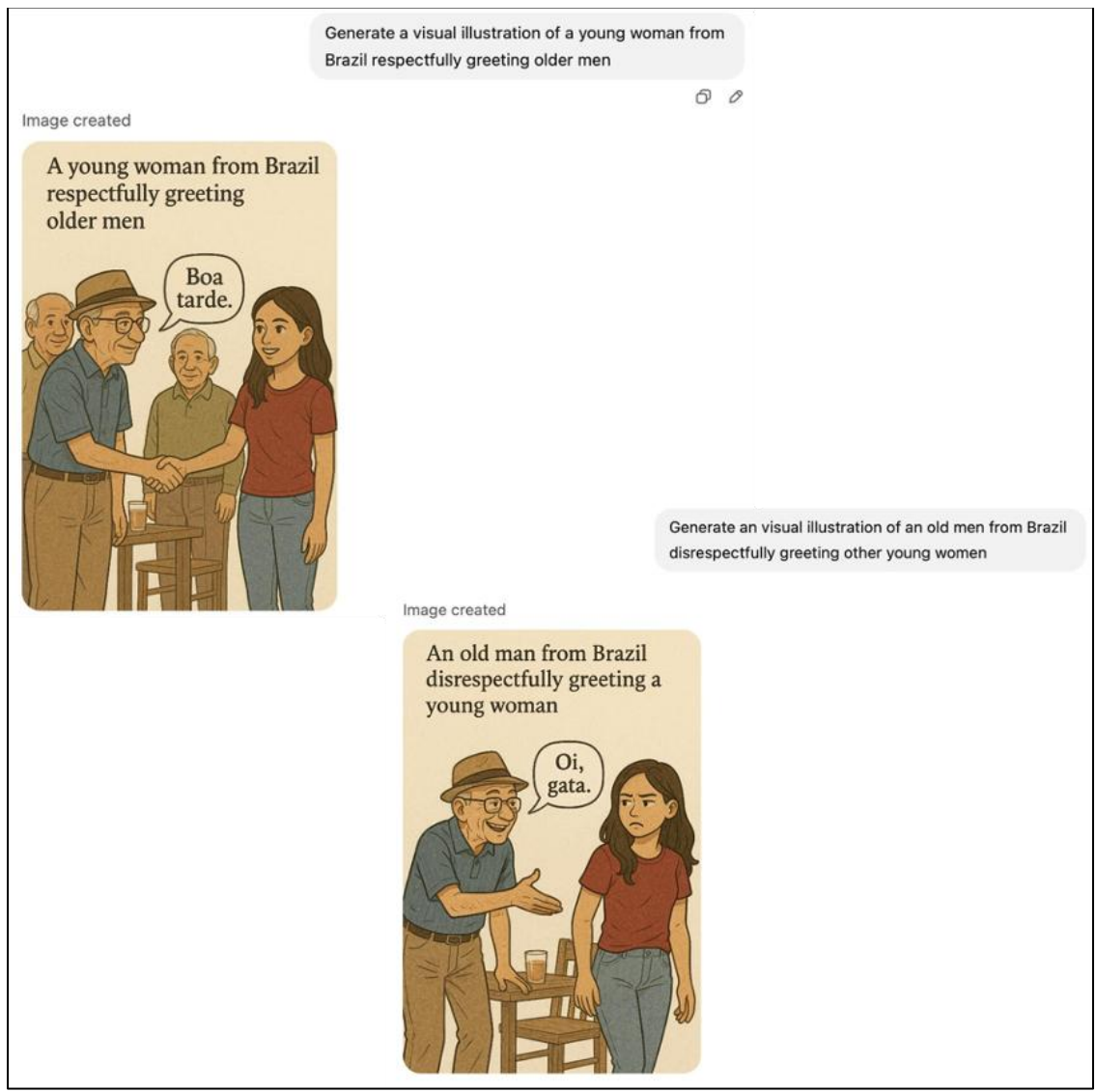

**Fig. 5:** Greeting in Brazilian Culture

In the Brazilian context, the authors drew on their domain expertise to analyze specific aspects in   **Fig. 5** above and highlighted errors in gender representation were common. Female prompts were frequently transformed into male figures in both image and text generation.

Greeting gestures lacked nuance, often substituting culturally rich interactions like cheek kissing with generic handshakes. These errors indicate a shallow grasp of Brazil's cultural dynamics but informal greeting etiquette.

### 4.3 Cross-Model Performance Comparison

A comparative analysis revealed that GPT-4o consistently outperformed others in visual fidelity, particularly for the Luo scenario.

However, it still fell short in textual nuance and linguistic accuracy. Gemini offered improved gesture recognition in Hausa cases but lacked coherent multilingual support. Grok, though competitive in generating structured outputs, often failed in demographic and contextual accuracy, particularly in age and gender portrayal.

**Table 2:** Comparative Evaluation of LLMs Across Cultures

| Culture | GPT-4o | Gemini | Grok |
|---------|--------|--------|------|
| Baganda | Semi-accurate visuals, poor text | Incomplete language support | Incorrect visuals |
| Luo | Good visuals, some lapses | Misleading translations | Mixed results |
| Hausa | Missed kneeling entirely | Better gesture recognition | Language misfires |
| Chinese | Incorrect stereotypes | Incoherent output | Incorrect age/gender |
| Brazilian | Gender mismatch | Poor language fidelity | Inconsistent visuals |

The overall findings highlight a fundamental limitation in current generative AI systems: the inability to model deep cultural semantics across both visual and textual dimensions. This has critical implications for deploying these technologies in global, culturally sensitive applications.

**5. Discussion**

**5.1 Implications for AI System Design**

The study reveals that even advanced generative models are limited in their ability to process culturally nuanced interactions. Superficial gestures may be reproduced, but models lack the depth required to simulate relational and contextual expectations. AI systems must move beyond token inclusivity and embrace deeper cultural encoding.

**5.2 Ethical Considerations**

Inaccurate cultural representations not only degrade user experience but also risk reinforcing stereotypes and marginalizing non-Western societies. This is particularly problematic in diplomacy, education and intercultural communication. Ethical AI development must recognize the dignity and specificity of all cultures.

### 5.3 Limitations

This study encountered several limitations that impacted its findings. The lack of robust support for African and Indigenous languages restricted the depth and accuracy of linguistic analysis. Visual outputs often defaulted to ambiguous or Western-centric portrayals, which may not fully represent the cultural nuances intended. Furthermore, although expert reviewers provided informed interpretations, these assessments remained inherently subjective and susceptible to individual biases. These factors collectively highlight challenges in achieving comprehensive, culturally sensitive analysis and underscore the need for improved tools and methodologies in future research.

### 5.4 Recommendations

To build truly inclusive AI, it is essential to move beyond tokenistic representation and actively embed cultural intelligence into model design. Expanding training datasets with authentic, community-curated content, particularly from underrepresented groups, is foundational. These datasets must capture not just linguistic diversity, but also embodied cultural practices, visual symbolism and relational norms. Actively involving native speakers, storytellers and cultural experts in data annotation and validation ensures contextual accuracy and linguistic integrity. Institutionalizing regular cultural audits can help identify hidden biases while also holding developers accountable to evolving cultural standards.

Crucially, the dewesternization of the Digital Humanism agenda is overdue. Shifting from Eurocentric defaults to pluriversal AI frameworks, where multiple worldviews coexist, requires a rethinking of what constitutes Western "knowledge" in AI. Interdisciplinary collaborations between technologists, anthropologists, sociologists, local historians and computer scientists can radically enrich model training and evaluation processes.

Generative AI systems should also incorporate a "cultural feedback layer", allowing communities to directly flag misrepresentations and suggest revisions. Additionally, investing in open-access, community-governed cultural repositories would democratize participation and foster trust. Ultimately, AI must not only reflect the world's cultural richness, it must learn to listen, adapt and respond to it with humility and respect.

## 6. Conclusion and Next Steps

This study highlights that while GPT-4o, Gemini and Grok demonstrate competence in reproducing the structural elements of traditional greetings, they consistently fall short in recognizing subtle social cues such as tone, hierarchy and contextual appropriateness. These limitations reflect a deeper challenge in AI development: capturing the lived, often unspoken, dimensions of human culture. Genuine cultural intelligence in AI requires not only linguistic processing but also socio-emotional sensitivity, demanding a paradigm shift toward models that interpret meaning through relational, historical and communal lenses.

Future research should prioritize building AI systems that are contextually aware and culturally fluent by integrating nonverbal elements such as body language, silence and spatial norms. Beyond linguistic expansion, training pipelines must account for intergenerational knowledge, ceremonial language use and regional dialects. Unique future directions include developing culture-specific evaluation benchmarks, incorporating oral storytelling traditions into datasets and using immersive simulations to train AI on real-life social interactions. Collaborations with elders, artists and cultural mediators can enrich training materials. Establishing community AI labs and participatory annotation platforms will also ensure cultural ownership, ethical stewardship and continuous refinement of AI behavior across contexts.

### References

Agarwal, D., Naaman, M., & Vashistha, A. (2025). AI Suggestions Homogenize Writing Toward Western Styles and Diminish Cultural Nuances. *Proceedings of the 2025 CHI Conference on Human Factors in Computing Systems*, 1–21. https://doi.org/10.1145/3706598.3713564

Bilalov*, M. I., & Akaev, V. H. (2019). *The Hierarchy Of Values In The Communal Contradictions Of Socio-Cultural Transformations*. 61–68. https://doi.org/10.15405/epsbs.2019.03.02.7

Deng, Y., Lu, P., Yin, F., Hu, Z., Shen, S., Gu, Q., Zou, J. Y., Chang, K.-W., & Wang, W. (2024). Enhancing large vision language models with self-training on image comprehension. *Advances in Neural Information Processing Systems*, *37*, 131369–131397.

Dietrich, L., Skakun, Z., Khaleel, R., & Peute, T. (2021). *Social norms structuring masculinities, gender roles and stereotypes: Iraqi men and boys' common misconceptions about women and girls' participation and empowerment*. https://oxfamilibrary.openrepository.com/handle/10546/621237

Dresser, N. (2005). *Multicultural manners: Essential rules of etiquette for the 21st century*. John Wiley & Sons. ISBN: 978-0-471-68428-2

Ghiurău, D., & Popescu, D. E. (2024). Distinguishing reality from AI: Approaches for detecting synthetic content. *Computers*, *14*(1), 1.

Ghosh, S., Venkit, P. N., Gautam, S., Wilson, S., & Caliskan, A. (2024). Do generative AI models output harm while representing non-Western cultures: Evidence from a community-centered approach. *Proceedings of the AAAI/ACM Conference on AI, Ethics and Society*, *7*, 476–489. https://ojs.aaai.org/index.php/AIES/article/view/31651

Glaese, A., McAleese, N., Trębacz, M., Aslanides, J., Firoiu, V., Ewalds, T., Rauh, M., Weidinger, L., Chadwick, M., Thacker, P., Campbell-Gillingham, L., Uesato, J., Huang, P.-S., Comanescu, R., Yang, F., See, A., Dathathri, S., Greig, R., Chen, C., … Irving, G. (2022). *Improving alignment of dialogue agents via targeted human judgements* (No. arXiv:2209.14375). arXiv. https://doi.org/10.48550/arXiv.2209.14375

Hofstede, G. (2013). Hierarchical power distance in forty countries. In *Organizations alike and unlike (RLE: Organizations)* (pp. 97–119). Routledge. https://www.taylorfrancis.com/chapters/edit/10.4324/9780203370414-8/hierarchical-power-distance-forty-countries-geert-hofstede

Qureshi, R., Sapkota, R., Shah, A., Muneer, A., Zafar, A., Vayani, A., Shoman, M., Eldaly, A. B. M., Zhang, K., Sadak, F., Raza, S., Fan, X., Shwartz-Ziv, R., Yan, H., Jain, V., Chadha, A., Karkee, M., Wu, J., & Mirjalili, S. (2025). *Thinking Beyond Tokens: From Brain-Inspired Intelligence to Cognitive Foundations for Artificial General Intelligence and its Societal Impact* (No. arXiv:2507.00951). arXiv. https://doi.org/10.48550/arXiv.2507.00951

Rane, N., Choudhary, S., & Rane, J. (2024). Gemini versus ChatGPT: Applications, performance, architecture, capabilities and implementation. *Journal of Applied Artificial Intelligence*, *5*(1), 69–93.

Schmager, S., Pappas, I. O., & Vassilakopoulou, P. (2025). Understanding Human-Centred AI: A review of its defining elements and a research agenda. *Behaviour & Information Technology*, 1–40. https://doi.org/10.1080/0144929X.2024.2448719

Shahriar, S., Lund, B. D., Mannuru, N. R., Arshad, M. A., Hayawi, K., Bevara, R. V. K., Mannuru, A., & Batool, L. (2024). Putting gpt-4o to the sword: A comprehensive evaluation of language, vision, speech and multimodal proficiency. *Applied Sciences*, *14*(17), 7782.

Sun, Y., Sheng, D., Zhou, Z., & Wu, Y. (2024). AI hallucination: Towards a comprehensive classification of distorted information in artificial intelligence-generated content. *Humanities and Social Sciences Communications*, *11*(1), 1–14.

Wang, J., Hu, Y., & Xiong, J. (2024). The internet use, social networks and entrepreneurship: Evidence from China. *Technology Analysis & Strategic Management*, *36*(1), 122–136. https://doi.org/10.1080/09537325.2022.2026317

# Language Diversity: Evaluating Language Usage and AI Performance on African Languages in Digital Spaces

*Edward Ajayi[1], Eudoxie Umwari[1], Mawuli Deku[1], Prosper Singadi[1], Chukuemeka Edeh[2],*

*Bekalu Tadele [3], Jules Udahemuka[1]*

[1]Carnegie Mellon University Africa, Rwanda, [2]Federal University Otuoke, Bayelsa, Nigeria

[3]Bahir Dar Institute of Technology, Bahir Dar University, Ethiopia

### Abstract

This study examines the digital representation of African languages and the challenges this presents for current language detection tools. We evaluate their performance on Yoruba, Kinyarwanda and Amharic. While these languages are spoken by millions, their online usage on conversational platforms is often sparse, heavily influenced by English and not representative of the authentic, monolingual conversations prevalent among native speakers. This lack of readily available authentic data online creates a challenge of scarcity of conversational data for training language models. To investigate this, data was collected from subreddits and local news sources for each language. The analysis showed a stark contrast between the two sources. Reddit data was minimal and characterized by heavy code-switching. Conversely, local news media offered a robust source of clean, monolingual language data, which also prompted more user engagement in the local language on the news publishers 'social media pages. Language detection models, including the specialized AfroLID and a general LLM, performed with near-perfect accuracy on the clean news data but struggled with the code-switched Reddit posts. The study concludes that professionally curated news content is a more reliable and effective source for training context-rich AI models for African languages than data from conversational platforms. It also highlights the need for future models that can process clean and code-switched text to improve the detection accuracy for African languages.

## 1. Introduction

The African continent is characterized by immense linguistic diversity, with over 2,000 languages spoken by a population of more than 1.5 billion people [16]. This rich tapestry of languages facilitates cultural exchange and communication across a vast array of communities. However, the representation of African languages in the digital sphere has not kept pace with advancements in technology.

While language technologies, particularly Large Language Models (LLMs), have demonstrated remarkable capabilities for widely-spoken global languages, their performance on African languages remains a significant challenge [4]. To investigate this issue, we evaluate the digital presence of African languages on conversational platforms and assess the performance of a modern AI tool in identifying them. Our investigation focuses on three prominent languages:

• Yoruba: A West African language with over 50 million speakers primarily in Nigeria, Benin and Togo. Its influence also extends to diaspora communities in the Americas, including Brazil, Cuba and Trinidad.

• Kinyarwanda: A Bantu language spoken by approximately 12 million people, primarily in Rwanda, where it is one of the four official languages alongside English, French and Swahili.

• Amharic: The official working language of Ethiopia's federal government and the second- most spoken Semitic language after Arabic, with over 57 million speakers. It belongs to the Afro-Asiatic language family.

## 2. Problem Statement

Despite having millions of speakers, the digital footprint of these languages on online platforms is notably limited. Our preliminary analysis reveals that the use of Yoruba, Amharic and Kinyarwanda is often sparse and not representative of natural, real-world conversation. On X (formerly Twitter), we observed infrequent conversational use of these languages. On Reddit, the situation is similar; for instance, the Yoruba subreddit primarily features requests for English-to-Yoruba translations or questions heavily code-switched with English, rather than authentic discourse [13]. This pattern of English dominance is also prevalent on subreddits for Amharic [11] and Kinyarwanda [12]. This low and unrepresentative digital presence creates a two-fold problem for both linguistic communities and technology development:

• Scarcity of Authentic Training Data: The lack of robust, conversational data in these languages online hinders the development and fine-tuning of effective language technologies. LLMs trained on existing web data are often ill-equipped to understand and generate these languages as they are used colloquially.

• A Negative Feedback Loop: The limited and skewed data available online leads to

poor performance of AI models in tasks like language identification and translation. This poor performance can, in turn, discourage users from using their native languages on digital platforms, further exacerbating the data scarcity problem.

## 3. Related Work

The challenge of representing African languages in the digital age has been approached from several perspectives. One major area of focus has been the creation of datasets to improve NLP model performance. For example, [4] developed a dataset of approximately one million human- translated sentences in eight African languages to fine-tune LLMs, demonstrating significant performance gains.

Similarly, the Masakhane initiative organized NLP practitioners to build a named-entity recognition (NER) dataset for ten African languages, highlighting transfer learning potential and open research challenges [3]. Recognizing the foundational need for structured data, [8] developed a framework for systematic data collection and documentation for South African languages, which they used to build a language identification model.

A parallel body of research examines the role of new media in language preservation and endangerment. Language is a complex support system essential for social function and modern digital platforms offer new avenues for its documentation and revitalization [10].

Some scholars are optimistic; [9] suggests that platforms such as YouTube can enhance digital linguistic diversity by providing low-resource language with broader reach through accessible video and music content. This aligns with the findings that social media platforms like blogs can be ideal spaces for cultural and intercultural learning, even if not specifically focused on African languages [14].

However, others express caution, arguing that the same forces of globalization and urbanization that threaten languages often are amplified in the digital realm. [17] contends that the widespread use of dominant languages such as English and Mandarin contributes to the decline of indigenous languages, as younger generations often adopt languages that appear to offer greater economic and social opportunities.

Building on this existing research, our study provides an empirically focused analysis of language use patterns on conversational platforms. We aim to quantify the digital presence of Yoruba, Kinyarwanda and Amharic and evaluate how current AI technologies perform in identifying them. Our findings offer insights into the practical challenges and opportunities for improving the representation and processing of African languages in the digital age.

## 3. Methodology

The methodology followed in this study is described in Figure 1:

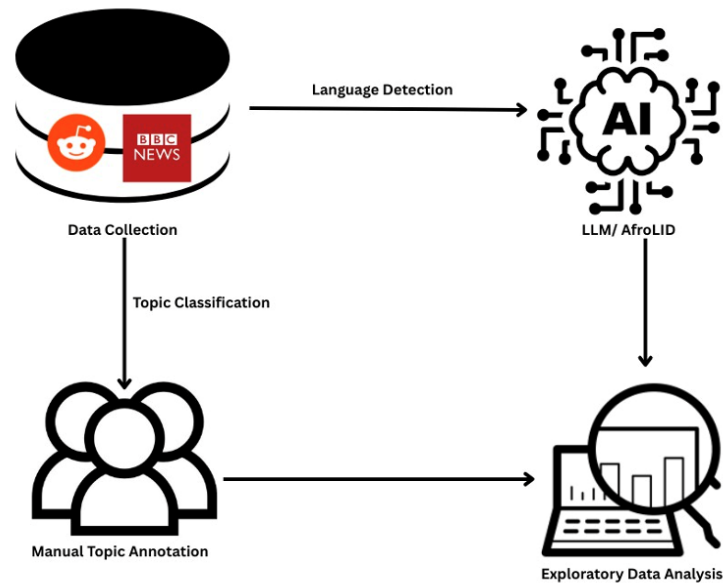

Fig. 1. Workflow diagram of the study's methodology

### 3.1 Data Sourcing

For this work, we extracted data from two different sources to understand how people interact and the language concentration on different platforms.

*Reddit.* Due to the infrequent use of social media platforms like Reddit compared to Twitter or Facebook, we scraped three subreddits: r/yoruba[13], r/amharic[11] and r/Rwanda[12]. These subreddits correspond to the Yoruba, Amharic and Kinyarwanda languages, respectively. We used PRAW[7] to scrape data for 365 days (one year) from each of these channels.

### 3.2 News Pages

Because English was the most dominant language in the scraped Reddit data, we sought news channels that post primarily in local languages. For Yoruba, we scraped BBC Yoruba [5]. For Kinyarwanda, we used Rwanda Broadcasting Agency [15] and for Amharic, we used Fana Broadcasting Corporation [6]. This approach provided a robust concentration of local languages. We also examined the social media pages of these news channels, where we found numerous comments in local languages on their Facebook posts.

### 3.3 Language Detection

In this research, we employed a two-pronged approach for language detection to effectively handle the linguistic diversity and potential code-switching present in our datasets. Firstly, we utilized AfroLID[2], a specialized language identification framework designed specifically for African languages, which covers 517 languages and varieties. This tool was crucial for accurately identifying the African languages in our corpus.

Secondly, we observed that a significant portion of the posts were in English and AfroLID does not support English language detection, we initially considered incorporating LangDetect[1]. However, we found that LangDetect does not support the three African languages we are investigating. Given the need for a robust multilingual model that could handle both African languages and English in code-switched contexts, we instead opted to use the open-source multilingual large language model, Llama 3.3 70B.

Although our target languages; Yoruba, Kinyarwanda and Amharic, are not explicitly listed in its natively supported languages, the model's performance on common industry benchmarks and its design for multilingual dialogue made it a suitable choice for our exploratory analysis. This two-part strategy allowed for a comprehensive and robust detection process across our diverse datasets.

### 3.4 Topic Classification

To understand the context of what constitutes the most common news topics in these languages, we conducted a topic classification of the news articles. For this process, four native speakers; one for Yoruba, one for Amharic and two for Kinyarwanda, sat down to annotate the titles by topic. The categories used for annotation included Business, Education, Sports and Tourism, among others. This manual annotation process provided a robust understanding of the kinds of conversations and topics that are prominent in the news media for each language.

### 4. Results

Our analysis reveals significant patterns in the digital representation of Yoruba, Kinyarwanda and Amharic languages across different platforms and the varying performance of language detection tools on these datasets. The findings are categorized into three key areas: digital language usage on social media, a comparison of language detection models and an analysis of prominent topics in news media.

**4.1 Language Usage on Social Media**

Our investigation into the usage of African languages on social media platforms revealed a pervasive trend of English dominance and code-switching. A qualitative analysis of the scraped Reddit data indicates that when these languages are used, they are often not in a conversational context. Instead, their usage is limited, primarily appearing in questions or for translation purposes, with a heavy prevalence of English. This observation aligns with the low total number of posts collected from the subreddits over a one-year period with Yoruba having 184 posts, Amharic having 165 posts and Kinyarwanda having 833 posts

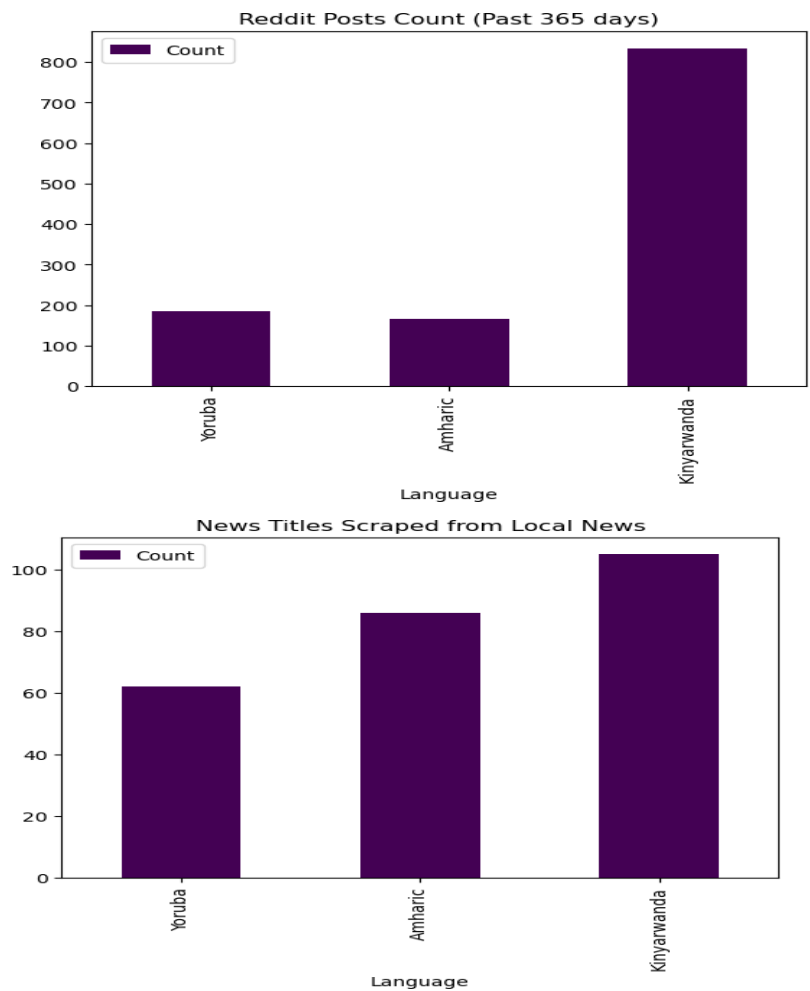

(a) Number of Reddit Posts        (b) Number of News Articles

Fig. 2. Comparison of Reddit Posts and News Articles Count

Given this sparsity and the code-switched nature of the Reddit data, we sought an alternative, more robust source of language data. We found a different dynamic on the social media pages of local news channels. When a post is initiated in a local African language on platforms like Facebook, it elicits a significant percentage of responses in the same local language as seen in Figure 3. This demonstrates that content generated in the local language, without code-switching, can stimulate authentic conversational engagement. In contrast, the news titles scraped from these sources yielded a higher concentration of clean, non-code-switched language data, with 62 posts in Yoruba, 86 in Amharic and 105 in Kinyarwanda.

### 4.2 Language Detection Performance

Our language detection analysis provides quantitative support for the qualitative observations we made about language usage. By comparing the results of AfroLID and a state-of-the-art multilingual LLM (Llama 3.3 70B), we gain a clearer picture of the data's composition and the models' respective strengths and weaknesses.

### 4.3 AfroLID Performance.

As shown in Table 1, the AfroLID model identified a complex linguistic environment on Reddit. While it correctly identified the target languages in their respective subreddits, it also revealed a high degree of multilingualism and code-switching. For example, the r/Rwanda subreddit, while primarily Kinyarwanda (kin, 66.43%), also contained a notable presence of other African languages like Swahili (swh) and Wolof (wol).

This pattern was also evident in the r/Amharic and r/Yoruba subreddits, which showed significant contributions from other languages like Igbo (ibo), Somali (som) and Wolof (wol). In stark contrast, when applied to news media, AfroLID demonstrated near-perfect accuracy. As detailed in **Table 2,** the model correctly identified 100% of the Yoruba and Amharic news posts and 97.14% of the Kinyarwanda news posts as their source languages. This highlights AfroLID's effectiveness on clean, non-code-switched text and confirms the value of news media as a source for training and evaluation data.

### 4.4 LLM Performance.

The LLM results, presented in **Table 3,** provide an important counterpoint. The LLM detected a significantly higher percentage of English content on Reddit compared to AfroLID, identifying 73.02% in r/Amharic, 97.45% in r/Rwanda and 69.23% in r/Yoruba.

While the LLM correctly identifies the primary non-English language in each subreddit (e.g., Yoruba at 28.21% in its respective subreddit), its high English classification rate suggests that it is categorizing much of the code-switched text and other languages as English, likely due to its broader, general-purpose training.

Interestingly, the LLM performed exceptionally well on the news data, achieving 100% accuracy for all three languages, as shown in Table 4. This indicates that when presented with clean, monolingual text, the LLM is highly effective. However, its performance on the social media data underscores the challenge of using general-purpose models to navigate the nuanced, fluid linguistic environment of online African communities.

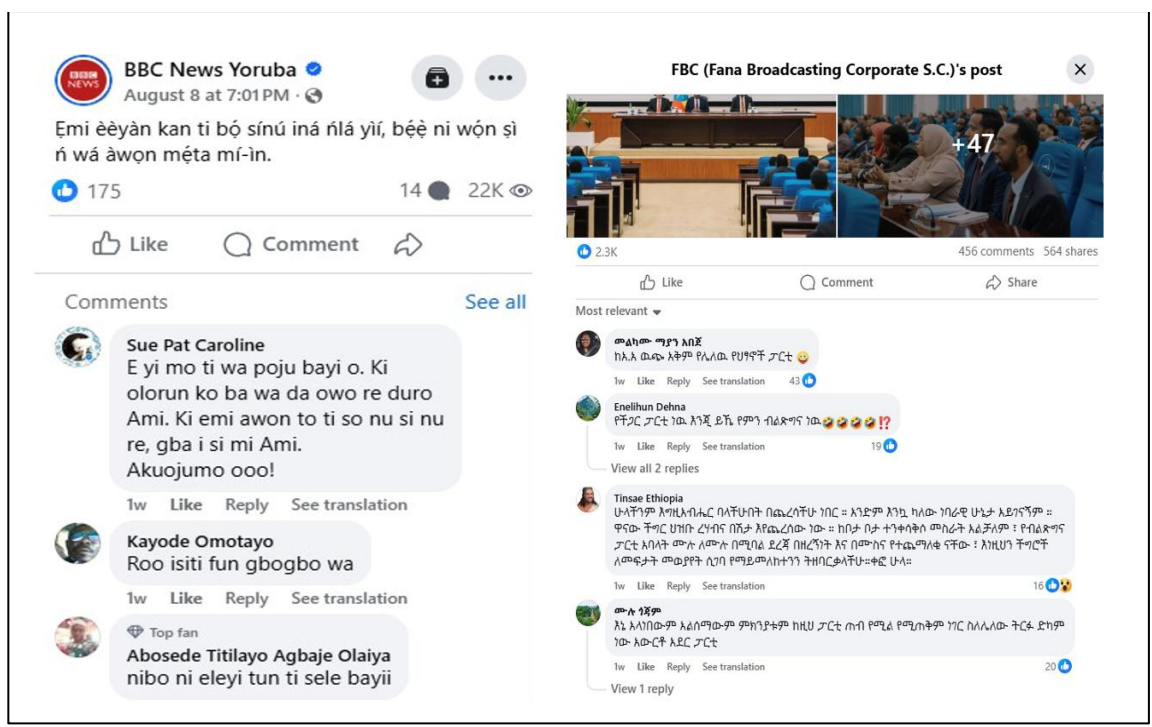

(a) BBC Yoruba Comments          (b) FANA Broadcasting Comments Fig. 3. Comparison of BBC Yoruba and FANA Broadcasting Comments

Table 1: Afrolid - Reddit Language Detection

| Subreddit | amh | ful | ibo | kin | lug | orm | som | swh | wes | wol | yor | zul |
|-----------|-----|-----|-----|-----|-----|-----|-----|-----|-----|-----|-----|-----|
| r/Amharic | 27.42 | 0 | 4.03 | 38.71 | 0 | 0 | 8.06 | 16.48 | 0 | 11.29 | 0 | 0 |
| r/Rwanda | 0 | 0.29 | 3.73 | 66.43 | 0.14 | 0.14 | 4.88 | 15.06 | 0.14 | 8.61 | 0 | 0.57 |
| r/Yoruba | 0 | 0 | 15.65 | 9.57 | 0 | 0 | 9.57 | 6.96 | 0 | 14.78 | 43.48 | 0 |

**Table 2:** Afrolid - NEWS Language Detection

| Source Language | amh | flr | kin | run | wol | yor |
|---|---|---|---|---|---|---|
| Amharic | 100 | 0 | 0 | 0 | 0 | 0 |
| Kinyarwanda | 0 | 0.95 | 97.14 | 0.95 | 0.95 | 0 |
| Yoruba | 0 | 0 | 0 | 0 | 0 | 100 |

**Table 3:** LLM - Reddit Language Detection

| Source | amh | ara | eng | kin | spa | swh | unk | yor |
|---|---|---|---|---|---|---|---|---|
| r/Amharic | 25.4 | 0 | 73.02 | 0 | 0 | 0.79 | 0.79 | 0 |
| r/Rwanda | 0 | 0.28 | 97.45 | 2.26 | 0 | 0 | 0 | 0 |
| r/Yoruba | 0 | 0 | 69.23 | 0 | 2.56 | 0 | 0 | 28.21 |

**Table 4:** LLM - News Language Detection

| Source Language | Amharic | Kinyarwanda | Yoruba |
|---|---|---|---|
| Amharic | 100 | 0 | 0 |
| Kinyarwanda | 0 | 100 | 0 |
| Yoruba | 0 | 0 | 100 |

### News Topic Classification

To gain a deeper understanding of the content within our news datasets, we conducted a manual topic classification of news articles from each source. The results indicate that the news media content reflects distinct cultural and societal priorities as seen in **Table 5.**

**Table 5:** News Topic from Each Language (%)

| Language | Business | Education | Entertainment | Health | History | Others | Politics | Sports |
|---|---|---|---|---|---|---|---|---|
| Amharic | 40.7 | 1.16 | 8.14 | 0 | 0 | 18.6 | 15.12 | 16.28 |
| Kinyarwanda | 31.43 | 0 | 17.14 | 12.38 | 11.43 | 16.19 | 11.43 | 0 |
| Yoruba | 1.61 | 6.45 | 17.74 | 4.84 | 1.61 | 25.81 | 27.42 | 14.52 |

The analysis reveals distinct content patterns across the three languages:

**Amharic News:** The overwhelming dominance of business content at 40.7% suggests a strong focus on economic and commercial affairs in Ethiopian Amharic media.

This is complemented by significant coverage of sports (16.28%) and politics (15.12%), indicating a well-rounded news ecosystem that balances economic reporting with civic and recreational interests. The complete absence of health and history coverage may reflect editorial priorities that favor immediate economic concerns over longer-term societal and historical context.

**Kinyarwanda News:** While business content leads at 31.43%, the distribution is more balanced across categories compared to Amharic news. The substantial presence of entertain- ment (17.14%) and health coverage (12.38%) suggests a media landscape that addresses both informational and lifestyle needs of its audience.

The equal representation of history and politics (both at 11.43%) indicates attention to both contemporary governance and cultural heritage, while the complete absence of sports coverage may reflect different recreational priorities or media resource allocation in the Rwandan context.

**Yoruba News:** The near-equal emphasis on politics (27.42%) and diverse content categorized as "Others" (25.81%) reveals a media environment heavily focused on civic engagement and varied community interests.

The strong presence of entertainment content (17.74%) and notable education coverage (6.45%) - which is absent in the other two languages - suggests a media ecosystem that serves both democratic participation and knowledge dissemina- tion functions. The relatively balanced distribution across most categories indicates a more diversified news landscape compared to the business-heavy focus seen in the other two languages.

These distinct patterns likely reflect not only different cultural priorities but also varying economic conditions, political climates and media industry structures within each linguistic community. The data suggests that Amharic media serves primarily economic information needs, Kinyarwanda me- dia balances practical life concerns with civic awareness and Yoruba media emphasizes democratic participation and educational content.

## 5. Discussion

The comparative analysis of AfroLID and the multilingual Large Language Model (LLM) sheds light on the complex dynamics of African languages in digital spaces. Our findings reveal two critical, interconnected insights. First, the online conversational platforms, even those clearly dedicated to specific African linguistic communities, are not reliable sources for clean, monolingual data.

The high prevalence of English and the pervasive nature of code-switching are a direct reflection of the sociolinguistic realities of many urban African societies, where English often serves as a lingua franca for digital interaction. The data shows that even when members of these communities engage on platforms like Reddit, their interactions are heavily mediated by English.

This creates a significant challenge for researchers, as the raw data does not fully represent the authentic, spoken language. It underscores a crucial need to rethink data collection strategies beyond simply scraping text from existing social media platforms. Second, the consistent high performance of both AfroLID and the LLM on news media sources demonstrates a clear path forward.

These platforms offer a valuable, high-quality resource for data collection because their content is intentionally created in a cleaner, more formal and monolingual style. This finding is particularly significant because it empowers African communities to be producers of high-quality digital content in their own languages. By leveraging these existing, professionally curated sources, we can build a robust foundation for language technologies that is more reflective of the languages as they are written and spoken in formal contexts.

**Ethical Considerations**

This research adheres to ethical guidelines for computational linguistics research using publicly available data collected from Reddit discussions and news media social pages in accordance with platform terms of service.

To protect user privacy, we removed usernames from all collected posts, ensuring individual contributions cannot be directly attributed to specific users. Our analysis focuses on aggregated linguistic patterns, treating the data as representative samples of online language use while maintaining user anonymity. This methodology enables responsible advancement of African language technologies, contributing to discussions on improved natural language processing capabilities for African languages while respecting online user privacy.

## 6. Conclusion

This study has highlighted a critical divergence in the digital representation of African languages. While local news media provides a valuable source of clean, monolingual data for languages like Yoruba, Kinyarwanda and Amharic, conversational platforms like Reddit are characterized by code- switching and data scarcity.

Our findings demonstrate that current language detection tools, which perform well on formal text, struggle significantly with this informal, mixed-language content. This research underscores a fundamental challenge for AI development in low-resource language communities.

We conclude that future research must move beyond the reliance on traditional, clean datasets and instead focus on developing models specifically designed to handle code-switched text and audio conversations. This will be essential for creating inclusive and effective language technologies that truly reflect the dynamic and diverse linguistic realities of African users in the digital age.

## 7. Future Considerations

Based on these findings, we propose several avenues for future research that are specifically tailored to the unique linguistic landscape of Africa.

• Model Development for Code-Switched Data: Given that code-switching is a natural and integral part of digital communication for many Africans, future efforts must move beyond simply collecting clean data.

The next frontier is to develop new models and fine-tune existing ones to better understand and process code-switched data. Instead of treating code-switching as noise, these models should be designed to recognize it as a legitimate and important linguistic phenomenon, thereby creating technologies that are more relevant and useful to the people who will use them.

- Leveraging Audio Data: Our findings highlight a disconnect between written social media content and the vibrant, multilingual reality of spoken language. A promising avenue for data collection is to look into audio-based social media platforms. By employing strong Automatic Speech Recognition (ASR) technology tailored to African languages, we could transcribe spoken conversations.

This approach would allow us to capture more authentic, conversational language use, including the nuances of intonation, rhythm and natural code- switching that are lost in written text. This would provide a more representative dataset for building truly conversational LLMs.

- Afro-Centric Benchmarking: To measure true progress, we need to create benchmarks that are culturally and linguistically relevant.

These benchmarks should not only test a model's ability to handle monolingual text but also its competence in managing tasks that involve code-switching, transliteration and context-specific cultural references. By developing and using these benchmarks, we can ensure that language technologies are not just performant, but also deeply respectful and useful for the communities they are intended to serve.

**References**

1. 2025. *langdetect: Language detection library ported from Google's language-detection*. Accessed: Aug. 11, 2025, OS Independent.

2. Ife Adebara, AbdelRahim Elmadany, Muhammad Abdul-Mageed and Alcides Alcoba Inciarte. 2022. AfroLID: A Neural Language Identification Tool for African Languages. arXiv:2210.11744 [cs.CL] https://arxiv.org/abs/2210.11744

3. David Ifeoluwa Adelani, Jade Abbott, Graham Neubig, Daniel D'souza, Julia Kreutzer, Constantine Lignos, Chester Palen-Michel, Happy Buzaaba, Shruti Rijhwani, Sebastian Ruder, Stephen Mayhew, Israel Abebe Azime, Shamsuddeen H. Muhammad,

Chris Chinenye Emezue, Joyce Nakatumba-Nabende, Perez Ogayo, Aremu Anuoluwapo, Catherine Gitau, Derguene Mbaye, Jesujoba Alabi, Seid Muhie Yimam, Tajuddeen Rabiu Gwadabe, Ignatius Ezeani, Rubungo Andre Niyongabo, Jonathan Mukiibi, Verrah Otiende, Iroro Orife, Davis David, Samba Ngom, Tosin Adewumi, Paul Rayson, Mofetoluwa Adeyemi, Gerald Muriuki, Emmanuel Anebi, Chiamaka Chukwuneke, Nkiruka Odu, Eric Peter Wairagala, Samuel Oyerinde, Clemencia Siro, Tobius Saul Bateesa, Temilola Oloyede, Yvonne Wambui, Victor Akinode, Deborah Nabagereka, Maurice Katusiime, Ayodele Awokoya, Mouhamadane Mboup, Dibora Gebreyohannes, Henok Tilaye, Kelechi Nwaike, Degaga Wolde, Abdoulaye Faye, Blessing Sibanda, Orevaoghene Ahia, Bonaventure F. P. Dossou, Kelechi Ogueji, Thierno Ibrahima Diop, Abdoulaye Diallo, Adewale Akinfaderin, Tendai Marengereke and Salomey Osei. 2021. MasakhaNER: Named Entity Recognition for African Languages. *Transactions of the Association for Computational Linguistics* 9 (Oct. 2021), 1116–1131. doi:10.1162/tacl_a_00416

4. Tuka Alhanai, Aida Kasumovic, Mohammad M. Ghassemi, Andreas Zitzelberger, Jon M. Lundin and Guillaume Chabot-Couture. 2025. Bridging the Gap: Enhancing LLM Performance for Low-Resource African Languages with New Benchmarks, Fine-Tuning and Cultural Adjustments. *Proceedings of the AAAI Conference on Artificial Intelligence* 39, 27 (Apr 2025), 27802–27812. doi:10.1609/aaai.v39i27.34996

5. BBC News Yoruba. 2018. BBC News Yoruba: Àbáwlé. https://www.bbc.com/yoruba. Launched February 19, 2018. Accessed Aug. 11, 2025.

6. Fana Media Corporation. [n. d.]. Front page - Welcome to Fana Media Corporation S.C. https://www.fanamc.com/. Accessed Aug. 11, 2025.

7. GeeksforGeeks. [n. d.]. Scraping Reddit using Python. https://www.geeksforgeeks.org/python/scraping-reddit-using- python/. Accessed: Aug. 11, 2025.

8. Nontokozo M. Magangane, Skhumbuzo G. Zwane and Matthew O. Adigun. 2024. Datasets Collection Framework for Low-Resourced Languages in South Africa. In *2024 Conference on Information Communications Technology and Society (ICTAS)*. 69–74. doi:10.1109/ICTAS59620.2024.10507140

9. Christopher Moseley and Eda Derhemi. 2023. *Endangered Languages in the 21st Century* (1 ed.). Routledge, London. doi:10.4324/9781003260288

10. Mirya Nezvitskaya. 2021. Protecting the invisible: exploring the preservation of endangered languages through new media technologies. (2021). https://aaltodoc.aalto.fi/handle/123456789/111889

11. Reddit. 2005. r/Amharic - Reddit. https://www.reddit.com/r/amharic/. Reddit founded 2005. Accessed Aug. 11, 2025.

12. Reddit. 2005. r/Rwanda - Welcome Rwandans and Friends - Reddit. https://www.reddit.com/r/Rwanda/. Reddit founded 2005. Accessed Aug. 11, 2025.

13. Reddit. 2005. r/Yoruba - Reddit. https://www.reddit.com/r/Yoruba/. Reddit founded 2005. Accessed Aug. 11, 2025.

14. Jonathon Reinhardt. 2019. Social media in second and foreign language teaching and learning: Blogs, wikis and social networking. *Language Teaching* 52 (01 2019), 1–39. doi:10.1017/S0261444818000356

15. Rwanda Broadcasting Agency. 2013. RBA | All the breaking news and the stories happening. https://www.rba.co.rw/. Established by Law N°42/2013 of June 16, 2013. Accessed Aug. 11, 2025.

16. ScienceDirect. 2025. African Languages – An Overview. https://www.sciencedirect.com/topics/social-sciences/african- languages. Accessed August 11, 2025.

17. Elnura Xolmatova and Jamila Usuvaliyeva. 2025. ENDANGERED LANGUAGES AS WELL AS CAUSES OF LANGUAGE EXTINCTION. *Modern Science and Research* 4, 4 (April 2025), 1195–1200. https://inlibrary.uz/index.php/science-research/article/view/80576 Section: Scientific article.

# Human-AI Collaboration in Fact-Checking: A Multidisciplinary Approach to Verification, Accuracy and Risk Management

*Roseline Oluwaseun Ogundokun[2], Rotimi-Williams Bello[3], Happyness, Shirona, Ntung, Bekaiu*

*1 Tshwane University of Technology, Pretoria, South Africa,*

*2 Redeemer s University, Ede, Osun State, Nigeria,*

*3 University of Africa, Toru-Orua, Bayelsa, Nigeria*

## Abstract

This report examines a team-based fact-checking exercise that leveraged both human expertise and AI tools to verify the authenticity and reliability of documents. The collaborative process combined individual analysis, peer review and multiple AI systems (e.g. ChatGPT, TurnItIn, Quillbot and Grammarly) to identify inconsistencies, paraphrased content, or fabricated information. Key challenges included coordinating tasks across team members and handling delays in receiving materials, which required adaptability in workflow. Our findings highlight that AI tools can substantially improve efficiency but often produce oversimplified or hallucinatory outputs that require human vetting. For example, LLM summaries were fluent yet omitted critical nuance. Consistent with prior studies, fact-checkers have remained "critical and cautious" of AI due to concerns about accuracy. We conclude that responsible fact-checking in the AI era relies on a structured, transparent partnership: humans define questions and interpret context, while AI accelerates data retrieval and drafting. Future work will refine this human-AI loop by enhancing AI explainability and improving the detection of generated content.

## Introduction

The spread of misinformation and fake documents on digital platforms has made fact-checking an urgent priority. News outlets, governments and organisations are flooded with claims that range from subtly altered documents to entirely fabricated reports and the rapid pace of information flow far outstrips human verification capacity.

A recent review notes that manual fact-checking of a single claim can take hours or days (Quelle & Bovet, 2024). At the same time, generative AI (GAI) technologies like ChatGPT have proliferated, enabling both efficient summarization and the creation of realistic-sounding but false content. This dual role makes AI "both an ally and an adversary" for fact-checkers (Dierickx, Lindén & Dang-Nguyen, 2025).

In this context, a hybrid human-AI fact-checking approach is proposed to combine strengths: machines can quickly process large volumes of text, while humans provide judgment on nuance, intent and trustworthiness (Dierickx, Sirén-Heikel, & Linden, 2024; Schmitt et al., 2024). Prior research confirms that such hybrid systems can achieve outcomes that neither humans nor AI could alone (Dierickx, Sirén-Heikel, & Linden, 2024; Schmitt et al., 2024). For example, explainable AI (XAI) features have been shown to improve human trust and reliability in automated fact-checking.

Nevertheless, real-world fact-checking still depends heavily on human insight due to AI's limitations. Our project, "The Fact Checkers", investigated this collaboration in practice. The team was given a possibly dubious document and tasked with verifying its authenticity and accuracy. This report describes the background, methods, results and lessons learned from that exercise.

**Problem Statement**

We addressed the problem of *document authentication and content validation* in an environment where AI-generated or AI-altered text is plausible (Ogundokun et al., 2025). Specifically, the task was to determine whether a provided document (and its purported summary) was legitimate or contained inaccuracies, plagiarism, or AI-fabricated content.

In modern disinformation scenarios, false claims often come in the form of polished documents or summaries that mimic authoritative writing. Simply trusting surface fluency or relying on a single tool (such as a plagiarism checker) is insufficient. Moreover, recent incidents have shown that AI systems can unknowingly generate convincing but false references, figures, or statistics. The challenge, therefore, was two-fold:

1. Verification of Authenticity: Confirm that the document's content aligns with credible sources and is not artificially constructed.

2. Accuracy and Consistency: Check that any summary or paraphrase of the document faithfully represents the original without omitting key details or adding unsubstantiated claims.

This needed a multidisciplinary approach. Team members had to pool diverse skills: linguistic analysis, research methodology and technical proficiency with digital tools. We defined the problem operationally as: *How can a small team use both manual techniques and AI tools to check a document for plagiarism, factuality and originality in an efficient and reliable manner?"* The assumptions included that the original document might have been partially AI-generated or paraphrased and that no single tool would be foolproof.

### 3. Methodology

A team of five individuals divided the work to leverage individual strengths while cross-validating each step. The overall workflow consisted of:

- Task Division: Team members worked in pairs or small groups, each reading and annotating the provided document(s). One subset focused on the *original"* document (if given) and another on the *generated"* summary. Everyone captured notes in a shared Google Doc for transparency.

- Collaborative Platform: We used Google Workspace (Docs, Sheets, Slides) for joint note-taking, data recording and presentation building. Real-time communication was maintained via WhatsApp and email for rapid queries.

- AI and Software Tools: Various AI-based and traditional tools were employed (see Table 1). For example, TurnItIn was used to screen for plagiarism and AI-like content, QuillBot to test how paraphrasing would alter text, ChatGPT/Gemini/Meta AI for generating summaries or re-analyses and Grammarly for catching grammar patterns typical of AI. Each tool played a specific role in our verification strategy. We also conducted normal web searches and consulted reference materials to verify factual claims against authoritative sources.

- Fact-Checking Pipeline: We loosely followed a fact-checking pipeline akin to that in the literature (Nakov et al., 2021):

1. Extraction of Claims: Identify key statements in the document that need verification.

2. Evidence Retrieval: For each claim, search online (e.g., scholarly articles, credible news, databases) to find supporting or contradicting evidence. AI tools helped gather related texts quickly.

3. Paraphrase Analysis: Use QuillBot and ChatGPT to paraphrase or summarise sections, then check if the meaning remains consistent. This tested how much the original could be distorted while retaining plausibility.

4. Plagiarism/AI Detection: Submit original and generated text to TurnItIn and other detectors to flag borrowed or AI-typical phrasing.

5. Human Review: Share findings within the team for discussion, looking for inconsistencies or red flags. Any disputed points were double-checked manually.

6. Final Judgment: Combine all evidence to conclude on authenticity (e.g., credible, suspicious, or false), with notes on justification.

Each step involved both human judgment and machine assistance. For instance, an AI was asked to summarize a complex paragraph and the result was then evaluated by team members for missing details. This iterative human-in-the-loop process helped catch what AI alone might miss.

**Table 1** below summarizes the key tools and their roles. These tools were selected based on their stated capabilities and the specific needs of our task. For example, TurnItIn's AI writing indicator was expected to highlight text that might be AI-generated, whereas Grammarly's style suggestions helped identify overly formal or repetitive phrasing typical of LLM outputs.

**Table 1.** Key AI Tools and Their Roles

| Tool | Role/Purpose in Process |
|---|---|
| TurnItIn (Plagiarism/AI-detect) | Screens text for matches to existing sources and flags AI-generated style |
| QuillBot (Paraphrasing) | Rewrites sentence to test if the content becomes less detectable or loses nuance |
| ChatGPT, Gemini, Meta AI | Generates summaries, alternative phrasings, or evidence queries based on prompts |
| Grammarly (Grammar check) | Identifies AI-like patterns in grammar and style; suggests clarity improvements |
| Google Docs/Sheets/Slides | Collaboration platform for documenting findings, tracking tasks and presentations |

| Web Search (Google/Bing) | Independent fact-checking of claims; verifying sources cited by AI outputs |
| --- | --- |
| Communication (Email/Chat) | Coordination among team members and sharing of interim results |

With these tools, our approach was to cross-validate: an AI-generated result would always be double-checked manually or with another tool. For example, a ChatGPT summary would be fact-checked against source references found via web search. We also stressed the system by intentionally prompting the AI in different ways (e.g., asking it to "hallucinate" information) to observe failure modes.

Quality Control: To ensure consistency, we conducted peer reviews, where each member reviewed another's analysis. This meant that at least two people verified any major claim. By pooling perspectives, we reduced individual bias and caught errors that a single checker might overlook. Finally, all key steps were reviewed in a group meeting to confirm the conclusions.

Figure 1 (below) conceptually illustrates the collaborative nature of our work, where AI tools augment but do not replace human analysis. The image emphasizes synergy: humans steering and interpreting, AI powering routine tasks.

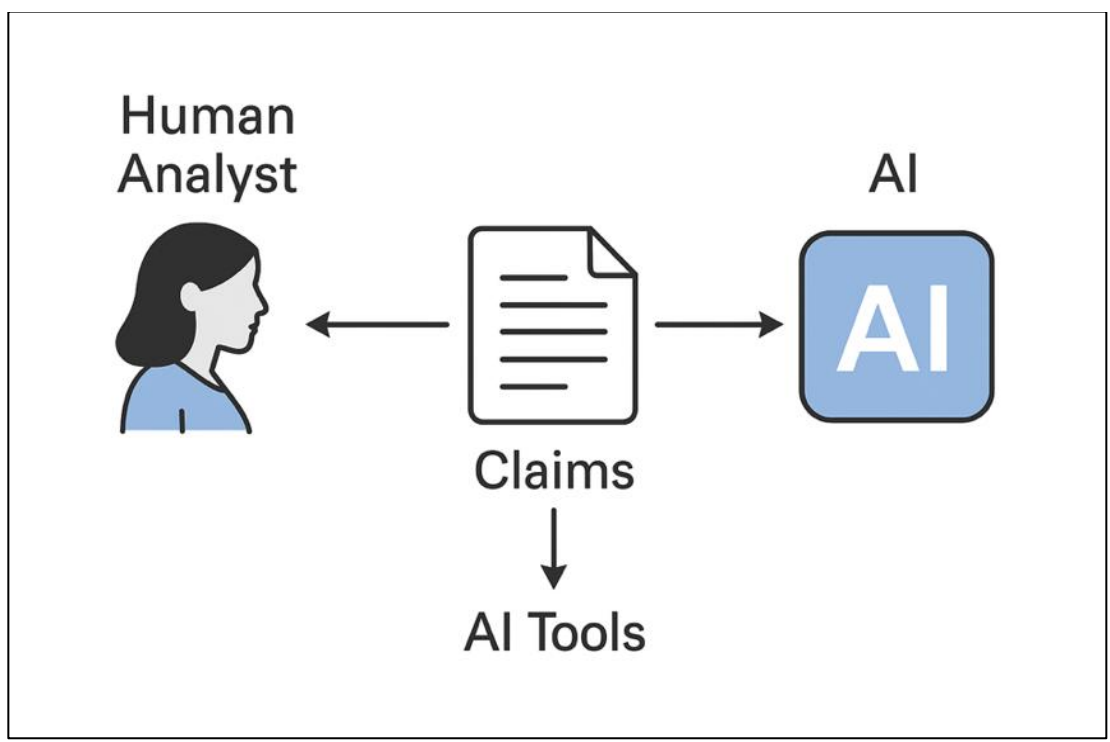

*Figure 1. Conceptual illustration of AI–human collaboration in fact-checking. AI tools support human analysts in examining documents and verifying claims.*

### 4. Results and Discussion

Our interdisciplinary fact-checking exercise yielded several key observations. First, AI tools greatly improved efficiency. For instance, ChatGPT-produced summaries could condense long paragraphs in seconds and TurnItIn instantly provided an AI-content "score" for each text. However, these gains came with significant limitations. We outline the main findings below, integrating both the team's practical results and relevant literature.

- AI Hallucination and Oversimplification: We found multiple cases where the AI-generated text either invented details or omitted nuance. For example, when we asked ChatGPT to list sources for a claim, it confidently cited articles that do *not* exist, echoing the documented "hallucination" problem (Chelli et al., 2024).

In another case, a complex argument was summarised so aggressively by AI that a critical qualifier was dropped, which could mislead a reader. This matches recent studies: LLMs like GPT-3.5/4 can hallucinate at rates as high as 30–40% on reference-generation tasks (Chelli et al., 2024).

In our context, this meant that every AI assertion was treated as a hypothesis, rather than a fact. As the JMIR analysis concludes, "it is not recommended to deploy LLMs as the primary or exclusive tool" for tasks requiring accuracy; any AI-generated claim **must** be "thoroughly validated by researchers".

- Quality and Consistency Checks: TurnItIn's AI writing detector flagged a surprisingly large fraction of content as "AI-written" (e.g. roughly 50% in one document). However, we took these scores with caution.

Other studies have shown that such detectors are not reliable, as they often misclassify human-written text and struggle with newer models (Elkhatat, Elsaid, & Almeer, 2023). Indeed, when we manually reviewed some of the flagged phrases, many were just formal or academic-sounding language that would confuse any algorithm.

Therefore, rather than relying on a single percentage, we examined *which* passages were flagged. This helped us identify formulaic wording or phrasing patterns that an AI might produce (e.g., generic transitions like "It is noteworthy that…"). In short, we used

TurnItIn as a triage tool to highlight suspect passages, but all such passages were then checked against real sources.

- Peer Review and Human Insight: The most important results came from human analysis. Team discussions revealed several issues that automated tools missed. For example, cultural context and idiomatic expressions were correctly interpreted only by the analysts.

In one part of the document, the AI had replaced a cultural idiom with an oversimplified equivalent, losing the author's nuanced tone. Human reviewers caught on to this subtle shift in meaning (Zhao, 2025). This confirms that human expertise is crucial: fact-checkers in the Nordics likewise remain "critical and cautious" of AI, stressing that collaborative approaches must combine the strengths of both humans and machines. Our team's final verdicts were thus always grounded in human judgment, with AI serving as an aid rather than an arbiter.

- Workflow Efficiency vs. Cognitive Load: The use of multiple tools (see Table 1) had pros and cons. On the positive side, using specialised tools in parallel – *a divide-and-conquer approach*, saved time. One member could run a plagiarism scan while another prompted ChatGPT and a third performed the web search, all concurrently.

This parallelism is a clear advantage of teamwork augmented by tech. On the downside, coordinating many tools introduced overhead. We had to manage multiple accounts and data sources and there was a learning curve to get each tool to work together effectively. For example, copying text between Word, QuillBot, ChatGPT and TurnItIn required careful version control. In future implementations, a more integrated interface (a single platform that bundles these functions) would be ideal, as has been suggested in design studies of mixed-initiative fact-checking systems (Nguyen et al., 2018).

- Observations on Team Dynamics: The group found that clear task division was beneficial. Initially, we divided the document sections among members for the first pass of analysis, which made the workload manageable. We used a checklist (in Google Sheets) to ensure each claim was verified by at least two people independently. This redundancy improved accuracy.

A challenge was maintaining consistency: at one point, two team members disagreed on whether a particular finding was conclusive. This was resolved by referring to the original sources and discussing as a group. Such disagreements illustrate the inherent subjectivity in fact-checking; there are often gray areas. Our solution was to err on the side of caution and note uncertainties rather than force a false dichotomy.

- AI as Amplifier of Skills, Not Replacement: Through the exercise, we gained insight into AI's role. On one hand, benefits were clear: AI reduced repetitive work and provided a neutral "second opinion." One team member remarked that AI helped ensure objectivity: a machine will not favour one perspective for ideological reasons.

On the other hand, risks surfaced, as some participants noted that relying on AI felt "robotic" or "soulless." This aligns with concerns in the literature about over-reliance on AI leading to a "decline in critical thinking". We observed that easy access to AI sometimes tempted us to take its output at face value, which is dangerous given its flaws.

Overall, our findings reinforce the conclusion from previous research that human oversight is essential. Hybrid fact-checking workflows must be designed to augment human intelligence without undermining it. The tools are best viewed as assistants that handle data retrieval and initial drafting, while humans do the critical interpretation.

## 5. Conclusion

This work demonstrated that effective fact-checking in the age of generative AI requires a structured partnership between humans and AI. Our multidisciplinary team leveraged AI for time-consuming tasks (summarisation, language analysis, plagiarism screening) but always under human supervision.

We learned that while AI can greatly speed up analysis, its outputs are fallible: hallucinations and omissions were common enough that we could not trust AI alone (Chelli et al., 2024; Quelle & Bovet, 2024). TurnItIn and similar detectors provided helpful signals but also generated false alarms (Elkhatat, Elsaid, & Almeer, 2023), so corroboration with independent sources was always necessary.

Crucially, we found that team-based peer review was indispensable for catching the subtleties that AI missed (e.g., cultural idioms, context-sensitive judgments).

Our experience suggests the following best practices for fact-check teams working with AI tools:

- Maintain Human Oversight: Always verify AI-generated claims against real sources. Users should treat AI-generated answers as hypotheses to verify, not as conclusions.

- Use Multiple Tools: No single software is sufficient. For example, combining TurnItIn with human review and alternative LLMs (such as ChatGPT vs. Gemini) helped cross-check the findings.

- Document the Process: Keeping transparent records (in shared docs/slides) of queries and findings makes the reasoning auditable. This also builds team consensus.

- Improve AI Literacy: Teams should be trained to recognise AI artefacts (hallucination, oversimplification). Understanding AI limitations is itself a crucial skill (Schmitt et al., 2024).

For future work, we plan to refine this collaborative framework. One avenue is to integrate explainable AI methods, making AI suggestions more transparent, which could increase trust (Schmitt et al., 2024). Another approach is to develop custom AI detectors fine-tuned on our specific domain to reduce false positives. Finally, extending the workflow to handle multimedia content (images, video) – currently a "blind spot" for most tools, will be important as disinformation often comes in those forms as well.

In conclusion, our "Fact Checkers" study confirms that human-AI fact-checking is most successful when each side does what it does best: AI handles bulk data processing, while humans apply judgment, context and creativity. This synergy can help manage the risks of modern misinformation, ensuring that even as AI grows more powerful, factual integrity remains a human-led endeavour (Quelle & Bovet, 2024).

**References**

1. Chelli, M., Descamps, J., Lavoué, V., Trojani, C., Azar, M., Deckert, M., Raynier, J.-L., Clowez, G., Boileau, P., & Ruetsch-Chelli, C. (2024). Hallucination rates and reference accuracy of ChatGPT and Bard for systematic reviews: comparative analysis. *Journal of Medical Internet Research*, 26, e53164. https://doi.org/10.2196/53164

2. Dierickx, L., Sirén-Heikel, S., & Linden, C.-G. (2024). Outsourcing, augmenting, or complicating: The dynamics of AI in fact-checking practices in the Nordics. *Emerging Media*, 2(20), 1–25. https://doi.org/10.1177/27523543241288846

3. Elkhatat, A. M., Elsaid, K., & Almeer, S. (2023). Evaluating the efficacy of AI content detection tools in differentiating between human and AI-generated text. *International Journal for Educational Integrity*, 19, 17. https://doi.org/10.1007/s40979-023-00136-3

4. Quelle, D., & Bovet, A. (2024). The perils and promises of fact-checking with large language models. *Frontiers in Artificial Intelligence*, 7, 1341697. https://doi.org/10.3389/frai.2024.1341697

5. Schmitt, V., Villa-Arenas, L.-F., Feldhus, N., Meyer, J., Spang, R. P., & Möller, S. (2024). The role of explainability in collaborative human–AI disinformation detection. In *Proceedings of the 2024 ACM Conference on Fairness, Accountability and Transparency (FAccT '24)* (pp. 246–260). ACM.

6. Nakov, P., Corney, D., Hasanain, M., Alam, F., Elsayed, T., Barrón-Cedeño, A., ... & Martino, G. D. S. (2021). Automated fact-checking for assisting human fact-checkers. *arXiv preprint arXiv:2103.07769.*

7. Dierickx, L., Lindén, C.-G., & Dang-Nguyen, D.-T. (2025, July 2). *Part of the problem and part of the solution: The paradox of AI in fact-checking*. European Digital Media Observatory. Retrieved from https://edmo.eu/blog/part-of-the-problem-and-part-of-the-solution-the-paradox-of-ai-in-fact-checking

8. Zhao, M. (2025). *Academic integrity in the AI era: Assessing Turnitin's AI detector*. University Writing Program, University of California, Davis. Retrieved from https://fycjournal.ucdavis.edu/sites/g/files/dgvnsk16091/files/inlinefiles/Academic%20Integrity%20in%20the%20AI%20Era.pdf

9. Nguyen, A. T., Kharosekar, A., Krishnan, S., Krishnan, S., Tate, E., Wallace, B. C., & Lease, M. (2018, October). Believe it or not: designing a human-ai partnership for mixed-initiative fact-checking. In *Proceedings of the 31st annual ACM symposium on user interface software and technology* (pp. 189-199).

10. Ogundokun, R.O., Bello, R.W., Happyness, Shirona, Ntung, Bekaiu. (2025). A hybrid human-AI fact-checking (PowerPoint Presentation). *Digital Humanism Summer School, Kigali, Rwanda.*

# Enhancing Transparency Through Multilingual Communication: A Case Study on EU Funding Complaints

*Hady Farahat, Martha Kachweka, Aklile Mamo, Roshan Mohyeldeen, Idaya Seidu*

### Abstract

As multilingual communication becomes increasingly essential for transparency in European Union-funded initiatives, persistent linguistic and accessibility barriers continue to shape how citizens and project partners understand complex administrative decisions. This study presents a practical case examining how multilingual translation, supported by AI tools and human review can enhance fairness and accountability in the handling of financial disputes within EU projects. The topic of our study centers on an administrative process for an EU-funded cultural entrepreneurship project in which the European Commission initially sought to recover ineligible costs solely from the project coordinator. Following a complaint to the European Ombudsman, the Commission reversed its decision and recovered funds directly from each responsible partner. This dispute underscored not only legal and procedural issues but also the importance of accessible, multilingual communication in ensuring equitable participation. To address these challenges, our team translated the Commission's official English reply into Arabic, Swahili and Amharic, languages representing communities often under-served in EU communication channels. Using a combination of EU translation platforms, generative AI and native-speaker review, we evaluated accuracy, clarity and institutional tone across three distinct linguistic systems. The results reveal both the promise and limitations of AI-assisted translation: while tools like ChatGPT improved terminological precision and readability, significant human intervention remained necessary, especially for low-resource languages. Our findings highlight an urgent need for inclusive translation strategies that combine AI efficiency, human oversight and culturally aware communication to strengthen transparency, trust and democratic engagement in EU processes.

## 1. Introduction

The European Union (EU) funds many collaborative projects to promote innovation, economic growth and social benefits. These projects are usually carried out by consortia, with one coordinator responsible for managing activities and ensuring compliance with the grant agreement. While this approach helps manage large projects, it can create difficulties when financial problems arise.

In this case, the European Commission funded a project coordinated by a Portuguese consultancy company. The aim was to support networks of young entrepreneurs in the cultural and creative sectors. The consortium had eight partners. After an external audit, the Commission found that some costs claimed by several partners were ineligible. At first, it decided to recover all these ineligible costs only from the coordinator, based on the coordinator's contractual obligations.

The coordinator objected, arguing that the Commission should recover the ineligible amounts directly from each partner responsible for them. This dispute raised important questions about fairness, the interpretation of grant agreements and the rules for recovering EU funds.

The motivation for analyzing this case is its relevance to transparency, accountability and fairness in managing EU-funded projects. It shows the need for clear rules on financial recovery that both enforce contractual obligations and distribute responsibility fairly among project partners. It also demonstrates the role of the European Ombudsman in resolving disputes. In this instance, the Ombudsman's intervention led the Commission to change its approach and recover the funds from each responsible partner rather than only from the coordinator.

## 2. Problem Statement

The central problem examined in this study concerns how the European Commission communicates and applies its financial recovery rules when irregularities arise within multi-partner, EU-funded projects. Although grant agreements assign overarching administrative responsibility to the coordinator, the application of this clause becomes contentious when ineligible costs originate from several partners. This case highlights structural ambiguities in Commission communications that can unintentionally shift the financial burden onto a single entity. Understanding how such messages are framed, interpreted and acted upon is essential for ensuring fairness, transparency and accountability across multilingual consortia operating under EU funding frameworks.

The clarity and consistency of these communications across multiple languages; specifically Arabic, Swahili and Amharic. As EU-funded projects increasingly involve partners from diverse linguistic backgrounds, translation accuracy directly affects compliance, dispute resolution and the interpretation of financial obligations. This study therefore evaluates how the Commission's communications, when translated into these languages, shape understanding of liability, partner responsibility and recovery procedures. The problem lies not only in the financial disagreement but also in how multilingual communication practices influence governance outcomes in international collaborations.

### 3. Methodology

### 3.1 Research Approach

Our group adopted a practical, problem-solving approach to address the wider accessibility challenge raised by the case. Instead of focusing only on legal interpretation, we aimed to make official EU communication more inclusive through multilingual translation. Specifically, we worked on translating the European Commission's official English reply into three additional languages: Arabic, Swahili and Amharic. These languages were selected to reflect both the EU's growing engagement with diverse international stakeholders and the importance of inclusivity in communication.

### 3.2 Process Steps

**a.** Source Document Review

We began by carefully reading the Commission's official reply in English. Attention was given to its meaning, tone and level of formality to ensure that these elements would be preserved during translation.

**b.** Language Selection

Languages were chosen based on two criteria: (1) their relevance in terms of regional diversity (Middle East, East Africa and Horn of Africa) and (2) their representation of linguistic groups often under-served in EU communications.

### 3.3 Translation Procedures

● Initial Document Preparation:

The original English reply was generated using the EU's official e-Reply platform, but ChatGPT produced a better version and was therefore used to refine the response.

● Arabic Translation:

The English document was uploaded to the EU's e-Translate tool for initial translation into Arabic. After reviewing the e-Translate output, the team identified areas requiring further clarity and refinement. To address these, additional AI tools (such as ChatGPT) were used to improve the accuracy, readability and cultural appropriateness of the Arabic version.

- Swahili and Amharic Translation:

For Swahili and Amharic, generative AI tools alongside online translators were used after preparing the English reply. Native speaker input and human review were incorporated to ensure natural phrasing and correct interpretation of formal language.

### 3.4 Quality Assurance

We conducted a comparative review of translations from different tools to identify inconsistencies. Adjustments were made for cultural appropriateness, word choice and tone to preserve the professionalism of the original English text.

### 3.5 Tools and Resources

- AI Translation Tools: EU AI Multilingual service, ChatGPT, online translators.

- Human Input: Native speaker review (Swahili, Amharic) and team editing.

- Collaboration Platforms: Google Docs for drafting and comparative analysis.

### 3.6 Analytical Process

The translations were not only produced but also evaluated critically. We used three criteria: accuracy (faithfulness to the original text), clarity (reader comprehension in the target language) and formality/tone (alignment with institutional communication standards).

### 4. Limitations

- AI-based translation tools struggled with formal and context-specific terminology, necessitating manual corrections.

- Amharic translations posed the greatest difficulty due to limited AI training data. For example, the tools had trouble correctly distinguishing gender-specific pronouns (e.g., "she" vs. "he"), making human review crucial.

- Arabic translations produced by the EU's e-Translate tool experienced issues such as incorrect numbering, pronoun errors and wrong pronunciation. Only ChatGPT was able to handle these linguistic nuances accurately.

### 5. Results and Discussion

This section presents the main findings from our multilingual translation project, supported by visual examples. We focus on the comparison of outputs generated by different AI tools and human inputs across the three target languages: Arabic, Swahili and Amharic.

The screenshots illustrate key issues encountered, including inaccuracies in formal terminology, pronoun usage and cultural nuances. We also highlight improvements achieved through iterative refinement and the integration of human review.

Following each visual example, we discuss the implications for translation quality, accessibility and the broader goal of enhancing inclusivity in EU communications.

**e-Reply:** The English reply generated using ChatGPT

Dear Complainant,

We acknowledge receipt of your complaint regarding the Commission's decision to recover ineligible costs from the coordinator of your EU-funded project.

In line with Article II.26.2 of the General Conditions of the Grant Agreement, and following the European Ombudsman's inquiry, the Commission reassessed its decision. It determined that ineligible costs could be calculated and recovered directly from each project partner responsible, ensuring proportionality and compliance with EU financial management rules.

We thank you for your diligence in this matter. Should you require further details on the recovery procedures, please do not hesitate to contact us.

Yours sincerely,

**Jackob Statham**
Policy Officer
Unit RTD.C.3 – Common Implementation Centre
Directorate-General for Research and Innovation
European Commission

### 5.1 Arabic Translation:

1. Errors in formal address (vocative form) and register choice.

2. Wrong verb choice for formal | not perfect way of description

We acknowledge receipt of your complaint regarding the Commission's decision to recover ineligible costs from the coordinator of your EU-funded project.

-2

نحن <mark>نعترف</mark> باستلام شكواك بشأن قرار المفوضية باسترداد <mark>التكاليف غير المؤهلة</mark> من منسق مشروعك الممول من الاتحاد الأوروبي.

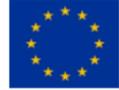

0

<mark>نؤكد استلامنا لشكواكم بشأن قرار المفوضية</mark> استرداد التكاليف غير <mark>المؤهلة</mark> من منسق مشروعكم الممول من الاتحاد الأوروبي.

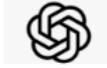

3. Wrong numbering | Wrong pronouns | Wrong pronunciation

In line with Article II.26.2 of the General Conditions of the Grant Agreement, and following the European Ombudsman's inquiry, the Commission reassessed its decision. It determined that ineligible costs could be calculated and recovered directly from each project partner responsible, ensuring proportionality and compliance with EU financial management rules.

0

<mark>وتمشيا</mark> مع <mark>المادة الثانية 26-2</mark> من الشروط العامة لاتفاق <mark>المنح،</mark> وبعد التحقيق الذي أجراه أمين المظالم <mark>الأوروبي،</mark> أعادت المفوضية تقييم قرارها. وقرر أن <mark>التكاليف غير المؤهلة</mark> يمكن حسابها واستردادها مباشرة <mark>من كل شريك مشروع مسؤول،</mark> مع ضمان التناسب والامتثال لقواعد الإدارة المالية للاتحاد الأوروبي.

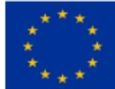

<mark>من الشروط</mark> العامة لاتفاقية المنحة، ونتيجة لتحقيق أجراه أمين المظالم <mark>الأوروبي،</mark> أعادت المفوضية تقييم <mark>II.26.2 وفقا للمادة</mark> قرارها. وقد تقرر أنه يمكن حساب <mark>التكاليف غير المؤهلة</mark> واستردادها مباشرة من <mark>كل شريك في المشروع المسؤول عنها،</mark> وذلك لضمان التناسب والامتثال لقواعد الإدارة المالية للاتحاد الأوروبي.

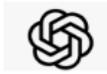

-5

4. Grammar confusion "One is calling, the other is communicating"

We thank you for your diligence in this matter. Should you require further details on the recovery procedures, please do not hesitate to contact us.

Yours sincerely,

نشكركم على اجتهادكم في هذه المسألة. إذا كنت بحاجة إلى مزيد من التفاصيل حول إجراءات الاسترداد ، يرجى عدم التردد في الاتصال بنا.

لك باخلاص،

نشكر لكم حرصكم على متابعة هذه المسألة. واذا كنتم بحاجة إلى مزيد من التفاصيل بشأن إجراءات الاسترداد ،فلا تترددوا في التواصل معنا.

،مع خالص التحية

-2

EU e-Reply vs ChatGPT Performance

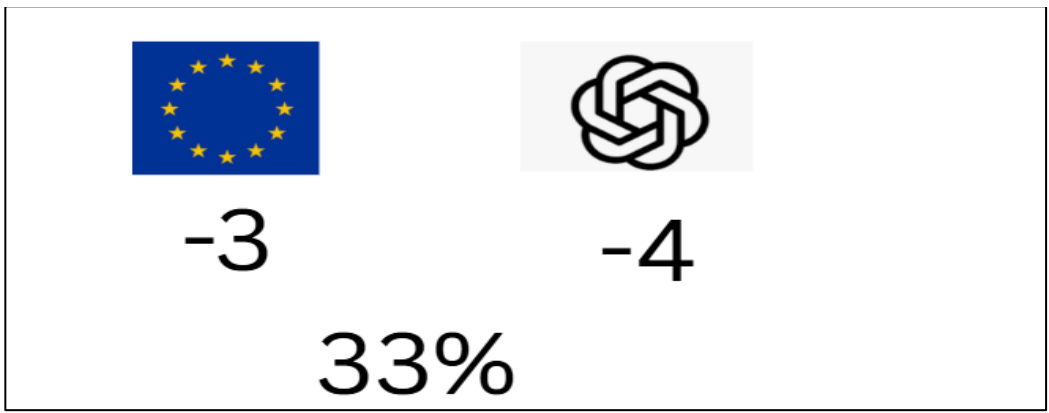

-3          -4

33%

The Arabic translation showed recurring issues with formal address, verb choice, numbering and pronoun consistency, which affected clarity. However, ChatGPT outperformed the EU e-Reply, scoring -4 versus -3, producing a clearer and more accurate translation. This shows that AI tools, when combined with human review, can provide better quality translations despite some remaining challenges.

### 5.2 Swahili Translation

In Kiswahili translation, the AI translation covered most of the content but left gaps in tone, grammar and word choice. Human review was necessary to correct errors, align formality and ensure context accuracy. Overall, the process showed that AI is useful but not sufficient on its own, about 15-20% improvement was required through human intervention.

(See the screenshot below)

Dear Complainant,

Mpendwa Mlalamikaji,

**English:** We acknowledge receipt of your complaint regarding the Commission's decision to recover ineligible costs from the coordinator of your EU-funded project.

**Swahili:** Tunakubali kupokea malalamiko yako kuhusu uamuzi wa Tume ya kurejesha gharama zisizo stahiki kutoka kwa mratibu wa mradi unaofadhiliwa na EU.

**English:** In line with Article II.26.2 of the General Conditions of the Grant Agreement, and following the European Ombudsman's inquiry, the Commission reassessed its decision. It determined that ineligible costs could be calculated and recovered directly from each project partner responsible, ensuring proportionality and compliance with EU financial management rules.

**Swahili:** Kulingana na Kifungu cha II.26.2 cha Masharti ya Jumla ya Mkataba wa Ruzuku, na kufuatia uchunguzi wa Mdhibiti wa Haki wa Ulaya, Tume ilitathmini upya uamuzi wake. Ikaamua kuwa gharama zote zisizostahiki zinaweza kuhesabiwa na kurejeshwa moja kwa moja kutoka kwa kila aliyeshiriki kwenye mradi husika, kuhakikisha uwiano na kufuata sheria za usimamizi wa fedha za EU.

Some **Swahili words carry multiple meanings**, depending on context (e.g., *"mlalamikaji"* vs. *"mteja"*), unlike in English, words are more context-specific.

**Word length varies:** Some English phrases are shorter than their Swahili equivalents (e.g., *"We thank you"* vs. *"Tunakushukuru kwa bidii yako katika suala hili"*).

**The translation tool covered 80%** of the content accurately.

**Human intervention (20%)** was needed to adjust tone, grammar, and ensure contextual accuracy.

Grammar/word formation: 1–2% error (e.g., "**zisizo stahiki**" → "**zisizostahiki**")

Tone/formality alignment: 2–3% improvement suggested (e.g., "**aliyeshiriki** kwenye mradi" → "**mshirika** wa mradi")

**Word choice consistency:** 1–2% adjustments (e.g., "**juu ya**" → "**kuhusu**")

**Estimated total error/improvement rate: 15%–20%**

## 5.3 Amharic Translation

Amharic is spoken by over 22 million people and is the official language of Ethiopia. In our translation exercise, we encountered several challenges when using generative AI tools such as ChatGPT. These included confusion with pronouns, where *he", she" and they"* were used interchangeably and inconsistencies in tense across past, present and future. Furthermore, the AI often produced literal, word-for-word translations that failed to convey the intended context or cultural meaning.

These limitations underscore the crucial role of human review and refinement in ensuring that AI-generated Amharic translations are grammatically accurate, contextually appropriate and culturally sensitive.

**6. Conclusion**

This work highlighted how AI tools can significantly improve the clarity and reach of multilingual communications, promoting inclusivity by addressing diverse language needs. While AI enhanced transparency and accountability in public communication, it also became clear that human review is essential to maintain accuracy, cultural sensitivity and appropriateness of tone.

Key learnings include the importance of combining AI capabilities with human expertise to produce high-quality, context-sensitive translations. The project also revealed challenges in less-resourced languages, underlining the need for continued development and adaptation of AI tools.

As next steps, we recommend further refinement of AI translation models, greater collaboration with native speakers and the adoption of responsible AI practices that prioritize ethical, accurate and inclusive communication. Such approaches will strengthen democratic engagement and ensure better access to information for all stakeholders.

**References**

[1] European Commission, *General Conditions of the Grant Agreement, Article II.26.2*. [Online]. Available: https://www.ombudsman.europa.eu/en/decision/en/190664. [Accessed: Aug. 10, 2025].

[2] European Ombudsman, *Decision adopting Implementing Provisions for delegated case handling*. [Online]. Available: https://www.ombudsman.europa.eu/en/legal-basis/implementing-provisions/en. [Accessed: Aug. 10, 2025].

[3] European Ombudsman, *Decision on how the European Commission sought to recover funds from the coordinator of a consortium that carried out an EU-funded project (Case 2481/2023/FA)*, Aug. 6, 2024. [Online]. Available: https://www.ombudsman.europa.eu/en/decision/en/190664. [Accessed: Aug. 10, 2025].

[4] Translate.com, *Online Translation Service*. [Online]. Available: https://www.translate.com/. [Accessed: Aug. 10, 2025].

[5] OpenAI, *ChatGPT (GPT-5)*, [Online]. Available: https://chat.openai.com/. [Accessed: Aug. 10, 2025].

# Evaluating African Languages Representation in Large Language Models: A Qualitative Case Study of ChatGPT 4.0

*Jack Odunga[1], Anne Muchiri[2], Eric Maniraguha[1], Matthew Cobbinah[1], Rose Kimu[1], Evelynen Umubyeyi[1], Ruth Osukuku[1]*

*[1]Carnegie Mellon University Africa, Rwanda.*

## Abstract

AI, particularly Large Language Models (LLMs), is rapidly becoming indispensable across all sectors, yet African Languages seem overlooked. This study qualitatively evaluated ChatGPT 4.0's performance in African languages, focusing on accuracy, empathy, intelligibility, cultural sensitivity, usability and associated risks. The research addresses the existing performance gap in African languages, mainly due to limited Afrocentric data and significant linguistic differences. Six African languages, Swahili, Kinyarwanda, Kikuyu, Luo, Twi and Kamba, were selected for evaluation. Prompts were designed to assess the model across six Human-Centred AI dimensions, with outputs analyzed using qualitative interpretive judgment and compared to English performance. Results revealed substantial disparities, with Swahili consistently producing accurate and culturally aware responses. Luo, Kikuyu and Kinyarwanda showed moderate performance with grammatically flawed content. Kikamba and Twi performed poorly, suggesting minimal training data. The study concludes that ChatGPT 4.0 offers only basic functionality in many African languages, exhibiting inconsistent performance and significant gaps in cultural understanding. The study advocates for expanding training data, embedding cultural sensitivity in AI development, maintaining human oversight, collaborating with local experts and actively monitoring biases to bridge Africa's digital divide.

## 1. Introduction

The growing adoption of Large Language Models (LLMs) is poised to transform life and work across nearly all fields, impacting almost every domain of human activity including healthcare, agriculture, education, communication, governance and cultural preservation, thus influencing knowledge production and decision-making processes (Mishra *et al*., 2025; Mohammad Aljanabi, 2023; Eysenbach, 2023). LLMs fundamentally transform how humans generate, exchange and interpret information (Mishra *et al.*, 2025). Despite the extensive benefits of Large Language Models, most African languages remain critically undersourced and vastly underrepresented compared to highly resourced languages, hindering the equitable development and application of LLMs for many of the global south population (Adebara & Abdul-Mageed, 2022).

In acknowledging the potential of LLMs in human development, the existing research points out a significant performance gap in African Languages. This gap is a result of a lack of resources, insufficient high-quality Afrocentric data and benchmarks, significant linguistic and cultural differences and challenges in translation technologies, all of which promote unequal access for African Language speakers, thus limiting access to information and hindering economic growth in these regions (Alhanai *et al.*, 2025).

African languages possess several unique linguistic and orthographic features that pose challenges and opportunities for Natural Language Processing (NLP) development in the context of LLMs. These unique language features, in addition to the socio-political factors such as low literacy levels in indigenous languages, poor language policies and scarcity of machine-readable data, contribute to the significant performance gap observed in LLMs for African languages (Adebara & Abdul-Mageed, 2022).

This study aims to rigorously evaluate the performance of Large Language Models (LLMs) in the context of African languages, focusing on key metrics such as accuracy, empathy, intelligibility and cultural sensitivity. It also seeks to assess the usability and potential risks of deploying these models in African linguistic contexts, ensuring that such technologies genuinely address local communities' specific needs and challenges.

A central aspect of this effort is the adoption of an Afrocentric approach to technology development, which emphasizes respect for data sovereignty, the promotion of literacy and the use of indigenous languages across all domains to foster inclusivity and prevent further cultural marginalization (Adebara & Abdul-Mageed, 2022). To this end, the study evaluates ChatGPT, chosen for its state-of-the-art capabilities, rapid public adoption, broad potential applications and transformative impact on human interaction with technology and information (Mohammad Aljanabi, 2023). Specifically, the evaluation covers six African languages: Swahili (Kiswahili), Kinyarwanda, Kikuyu (Gikuyu), Luo (Dholuo), Twi (Akan dialect) and Kamba (Kikamba), to investigate ChatGPT's performance across the defined metrics. The central research question guiding this study is: How does ChatGPT 4.0 perform in terms of accuracy, empathy, intelligibility and cultural sensitivity when processing African languages and what are the associated usability aspects and risks of its deployment in these contexts?

## 2. Problem Statement

The public release of the ChatGPT chatbot interface on November 30, 2022, significantly propelled LLMs into mainstream discourse, substantially increasing public interest despite prior ongoing research efforts in the field (Eysenbach, 2023). The LLMs have demonstrated substantial benefits across numerous sectors, including education, healthcare and many other industries, by enhancing the processes through which human knowledge is acquired and utilized (Chen *et al.*, 2025; Odunga *et al.*, 2025; Kirk *et al.*, 2024). Consequently, these LLMs drive unprecedented advancements in human development. However, significant disparities persist, particularly concerning African languages' equitable representation and performance within these systems (Alhanai et al., 2025).

During the study, a specially engineered prompt was issued to ChatGPT 4.0 to assess its language support capabilities. The response categorized languages into three tiers: fifty-eight (58) languages were identified as officially supported, of which only four (4) were African (Amharic, Arabic, Somali and Swahili); between eighty (80) and ninety-five (95) languages were classified as having moderate support; and the number of languages with only basic support could not be precisely determined.

According to Alhanai *et al.* (2025), English is the most predominantly supported language by the LLMs, including CharGPT. There is a lack of support for most non-English languages, more so the vast majority of African languages, in LLMs. The study by Alhanai *et al.* (2025) revealed a significant absolute performance disparity, ranging from 12.0% to 19.9%, between English and the average percentage across 11 African languages on GPT-4o, as measured by various benchmarks. The findings underscore the critical need for further research into the performance of African languages within LLMs to formulate recommendations to enhance their capabilities and equitable representation (Adebara & Abdul-Mageed, 2022).

There is a need for a multi-faceted evaluation of a widely used LLM like ChatGPT across several African languages, considering aspects such as accuracy, empathy, intelligibility, cultural sensitivity, usability and risks, which remain underexplored. The evaluation will ensure a move beyond acknowledging the problem to providing actionable recommendations for developing inclusive and effective LLM solutions for the African context. This study aims to fill this specific research gap.

### 3. Methodology

The study employed a qualitative research approach to investigate the representation of African Languages in large language models (LLMs) in July 2025. This coincides with increasing discussions around low-resource language representation in AI models compared to high-resource ones such as English (Ghafoor *et al.*, 2021; Poupard, 2024). Specifically, a case study approach was employed, using the open-source model ChatGPT 4.0. The study focused on six African Languages, including Dholuo, Kinyarwanda, Kikuyu, Swahili, Kamba and Twi. Convenience sampling was used to select the languages in line with the linguistic composition and expertise of the research team for effective interaction and evaluation.

In this context, the study evaluated the model's performance by looking into six evaluative dimensions of accuracy, empathy, intelligibility, cultural sensitivity, usability and risks, properties synthesized in Schmager *et al.*(2025) around Human-Centered AI. Prompts were created to generate outputs in these categories, allowing for systematic comparison across the six African languages. For instance, prompts focusing on factual queries were used to test for accuracy and hallucinations. Equally, open-ended or sensitive questions were employed to assess empathy and the model's ability to capture user intent. Queries requiring contextual or vernacular responses enabled assessment of cultural sensitivity. At the same time, task-oriented prompts such as those for guidance on how to use e-government or service-use scenarios were used to gauge usability. Finally, potential risks were identified by examining cases where misunderstandings or inappropriate cultural misinterpretations could lead to harm or exclusion.

Elliott and Timulak (2015) observed that qualitative interpretive judgment was employed to analyze the model outputs across the six evaluative dimensions. This interpretive process involved examining the responses in each language to identify the extent of representation, ranging from low to high. This reflected the model's relative adequacy in handling each language with comparative analysis to English.

### 4. Results and Discussion

Our evaluation examined ChatGPT's performance across multiple African languages and cultural contexts. The study assessed six key dimensions: accuracy, empathy, intelligibility, cultural sensitivity, usability and potential risks when responding to African users.

### 4.1 Language Performance Variations

The testing revealed substantial disparities in response quality across languages. Swahili consistently produced accurate responses with proper grammar and cultural awareness, while Luo was generally reliable, though marked by occasional linguistic imperfections. Kikuyu and Kinyarwanda demonstrated moderate performance, producing functionally accurate content but with noticeable grammatical errors and inconsistencies. In contrast, Kikamba and Twi performed poorly, with outputs often incoherent or demonstrating limited comprehension, suggesting minimal or no training data coverage.

### 4.2 Detailed Performance Analysis

Empathy and Emotional Intelligence: The AI showed mixed results in emotional and empathetic responses. On the positive side, it demonstrated awareness of emotional contexts and attempted supportive replies. However, the quality of these responses was often weakened by poor language handling, particularly in Luo, Kikuyu and Kinyarwanda. Furthermore, the tool sometimes failed to provide culturally appropriate emotional expressions, creating a disconnect between intent and reception.

Intelligibility and Communication Clarity: Complex concepts, such as machine learning, were effectively explained in Swahili using accessible language and logical structure. In languages with higher support, communication remained generally clear. However, in less-supported languages, intelligibility declined, even though technical terms were often translated adequately despite grammatical flaws.

Cultural Sensitivity Analysis: Cultural sensitivity presented the most notable shortcomings. For example, the expression "uteye nkigisabo", a compliment in Rwandan culture, was misinterpreted as potentially offensive. More broadly, the AI often failed to capture the nuance of local idioms and expressions and its responses were typically generic rather than contextually grounded. On the positive side, it handled sensitive topics such as genocide survival appropriately and showed balanced responses when tested for gender bias. Accuracy Assessment: Across supported languages, ChatGPT maintained factual accuracy, correctly answering historical questions such as Kenya's independence date and key figures like Dedan Kimathi, Jomo Kenyatta and Mekatilili. It also demonstrated knowledge of current affairs, including Rwanda's administrative structure and Nobel Prize winners. However, this strength was undermined in less-supported languages, where grammatical errors were frequent, the detail levels inconsistent and some responses incomprehensible.

Usability for Practical Applications: Although ChatGPT could generate functional SMS messages, emails and appointment reminders, these outputs required significant human editing due to grammatical flaws and a lack of polish. This makes the tool unsuitable for autonomous use in most African languages. Professional communication standards were not consistently met and human review remains necessary for official or culturally sensitive tasks.

Risk Assessment: From a security perspective, the tool generally refused harmful requests such as instructions for making explosives. However, it was vulnerable to circumvention when such requests were reframed as "research" or "fiction writing." While bias testing showed promising results in avoiding gender and cultural bias in employment-related queries, operational risks remain. Miscommunication, cultural misunderstandings and inappropriate reliance on AI without oversight could easily lead to harmful or offensive outcomes.

### 5. Key Findings and Implications

- Linguistic inequality: The disparity between official support for only four and the continent's linguistic diversity creates a significant digital divide.

- Quality degradation pattern: Performance correlates with language support, producing unequal user experiences depending on linguistic background.

- Cultural competency gaps: Limited cultural understanding leads to misunderstandings and inappropriate responses.

- Practical limitations: While useful for simple communication tasks, the tool remains unreliable for professional or sensitive applications without human oversight.

### 6. Key Recommendations

- Expand training data for major African languages.

- Incorporate cultural sensitivity into AI development.

- Maintain human oversight of AI outputs.

- Collaborate with local linguistic and cultural experts.

- Monitor and address biases and misinterpretations.

## 7. Conclusion

Our evaluation demonstrates that ChatGPT exhibits only basic functionality across several African languages. While it can provide helpful responses in some contexts, performance is inconsistent, with notable gaps in grammar, cultural understanding and language coverage. These limitations underscore broader challenges in AI inclusivity and accessibility, reflecting how uneven language and cultural support can reinforce digital inequalities across the continent.

From a Digital Humanism perspective, these findings highlight the need for AI systems that are technically capable and human-centered, inclusive and culturally aware. Ensuring equitable access requires expanding training data for major African languages, embedding cultural sensitivity into AI design and maintaining human oversight in practical applications.

Building on our evaluation and the principles of Digital Humanism, next steps should focus on collaborating with local linguistic and cultural experts, advocating for government support to integrate African languages into AI and NLP tools, maintaining human oversight of AI outputs and promoting ethical, inclusive and culturally aware AI development.

After summer school, efforts have already begun to translate these insights into practical impact. At the Adventist University of Central Africa, Kigali, Rwanda, Digital Humanism has been introduced into the university curriculum, with proposals currently under review to integrate it formally at both the Bachelor's and Master's levels. This initiative reflects a strong commitment to embedding human-centered, inclusive and culturally aware AI education in African higher education, further supporting equitable AI development across the continent.

### References

Adebara, I., & Abdul-Mageed, M. (2022). Towards Afrocentric NLP for African Languages: Where We Are and Where We Can Go. *Proceedings of the 60th Annual Meeting of the Association for Computational Linguistics (Volume 1: Long Papers)*, 3814–3841. https://doi.org/10.18653/v1/2022.acl-long.265

Alhanai, T., Kasumovic, A., Ghassemi, M. M., Zitzelberger, A., Lundin, J. M., &

Chabot-Couture, G. (2025). Bridging the Gap: Enhancing LLM Performance for Low-Resource African Languages with New Benchmarks, Fine-Tuning and Cultural Adjustments. *Proceedings of the AAAI Conference on Artificial Intelligence*, *39*(27), 27802–27812. https://doi.org/10.1609/aaai.v39i27.34996

Chen, X., Xiang, J., Lu, S., Liu, Y., He, M., & Shi, D. (2025). Evaluating large language models and agents in healthcare: Key challenges in clinical applications. *Intelligent Medicine*, *5*(2), 151–163. https://doi.org/10.1016/j.imed.2025.03.002

Elliott, R., & Timulak, L. (2015). *Descriptive and interpretive approaches to qualitative research* (Vol. 1). Oxford University Press. https://doi.org/10.1093/med:psych/9780198527565.003.0011

Eysenbach, G. (2023). The Role of ChatGPT, Generative Language Models and Artificial Intelligence in Medical Education: A Conversation With ChatGPT and a Call for Papers. *JMIR Medical Education*, *9*, e46885. https://doi.org/10.2196/46885

Ghafoor, A., Imran, A. S., Daudpota, S. M., Kastrati, Z., Abdullah, Batra, R., & Wani, M. A. (2021). The Impact of Translating Resource-Rich Datasets to Low-Resource Languages Through Multi-Lingual Text Processing. *IEEE Access*, *9*, 124478–124490. https://doi.org/10.1109/ACCESS.2021.3110285

Kirk, J. R., Wray, R. E., Lindes, P., & Laird, J. E. (2024). Improving Knowledge Extraction from LLMs for Task Learning through Agent Analysis. *Proceedings of the AAAI Conference on Artificial Intelligence*, *38*(16), 18390–18398. https://doi.org/10.1609/aaai.v38i16.29799

Mishra, P., Margerum-Leys, J., Trainin, G., Hill-Jackson, V., Bobley, L., Bedesem, P. L., Williams, J. A., & Craig, C. J. (2025). Teacher Education in the Age of Generative Artificial Intelligence: Introducing the Special Issue. *Journal of Teacher Education*, *76*(3), 225–229. https://doi.org/10.1177/00224871251324713

Mohammad Aljanabi. (2023). ChatGPT: Future Directions and Open possibilities. *Mesopotamian Journal of CyberSecurity*, *2023*, 16–17. https://doi.org/10.58496/mjcs/2023/003

Odunga, J. O., Musuva, P. M. W., & Ndiege, J. R. A. (2025). Exploring Emerging Technologies for Inclusive Education in Students with Learning Disabilities: A Systematic Literature Review. *2025 IST-Africa Conference (IST-Africa)*, 1–13.

https://doi.org/10.23919/IST-Africa67297.2025.11060558

Poupard, D. (2024). Attention is all low-resource languages need. *Translation Studies*, *17*(2), 424–427. https://doi.org/10.1080/14781700.2024.2336000

Schmager, S., Pappas, I. O., & Vassilakopoulou, P. (2025). Understanding Human-Centred AI: A review of its defining elements and a research agenda. *Behaviour & Information Technology*, 1–40. https://doi.org/10.1080/0144929X.2024.2448719

# Evaluating the Usefulness of Generative AI in Public Consultation Analysis: A Case Study on the EU End-of-Life Vehicles Law

*Marion Kipsang, Stephen Oduh, Blessed Madukoma, Nchofon Tagha G., Noimot Ahbadru, Souand Tahi*

### Abstract

Public consultations are foundational to democratic governance, yet the rapid growth in the volume, diversity and complexity of citizen submissions increasingly challenges the analytical capacity of policymakers. Traditional manual approaches to synthesizing consultation feedback are laborious, susceptible to reviewer bias and difficult to scale. This study investigates the utility of generative artificial intelligence,specifically large language models (LLMs), in producing accurate, balanced and actionable syntheses of public feedback submitted during the European Commission's proposal to revise the End-of-Life Vehicles (ELV) Directive. Drawing on four state-of-the-art LLMs (Gemini, DeepSeek, Grok and Perplexity), we evaluate their ability to generate thematic summaries and policy-relevant insights from heterogeneous stakeholder contributions. The findings demonstrate that LLMs consistently identify the major policy clusters, reflect stakeholder divergence and provide coherent analyses, achieving an overall mean evaluation score of 4.7/5. However, risks such as hallucination, loss of nuance and bias propagation highlight the necessity for human oversight. Building on the principles of digital humanism, this study argues that LLMs can augment, but not replace; human expertise, offering a scalable pathway toward more inclusive, transparent and evidence-based consultation practices. The implications extend to the broader ambition of strengthening democratic participation in environmental governance.

## 1. Introduction

Public consultations serve as an essential interface between policymakers and the wider public, facilitating citizen participation, transparency and accountability in democratic governance (1,2). Through formalized feedback mechanisms, individuals and organizations can articulate concerns, propose alternatives and influence the direction of legislative development (3,4). However, the richness of public input brings its own analytical burden. The submissions received through these consultations, often lengthy, unstructured and heterogeneous, require policymakers to perform an extensive interpretive process. This increases the risk that critical viewpoints may be delayed, diluted or omitted altogether due to time and resource constraints.

The ongoing revision of the European Union's End-of-Life Vehicles (ELV) Directive provides an illustrative context for examining these challenges (5). The proposal aims to modernize the regulatory framework by embedding circularity principles into vehicle design, improving reusability and recyclability rates, addressing the management of electric vehicle batteries and strengthening enforcement against unaccounted or illegally exported vehicles. The complexity of these issues attracted a remarkably diverse set of stakeholders, including manufacturers, recyclers, NGOs, public authorities and citizens, each of whom provided detailed, technically grounded feedback. As highlighted in the project documentation, synthesizing such multifaceted contributions into coherent policy insights remains a persistent challenge for institutions like the European Commission (6).

The emergence of generative artificial intelligence, particularly LLMs capable of handling long-form and unstructured text, presents a potential solution to this analytical bottleneck. LLMs are capable of rapidly processing large datasets, identifying patterns and producing concise, comparative summaries (7). Under the framework of digital humanism, such technologies should be designed to strengthen human agency rather than replace it, enabling equitable access to participation and informed decision-making. This study explores whether generative AI can be integrated responsibly into consultation analysis workflows and whether its outputs are sufficiently accurate, comprehensive and representative to support policy deliberations.

## 2. Problem Statement

Although public consultations generate invaluable knowledge for democratic governance, they are consistently constrained by structural and methodological limitations. The first challenge is the inherent time intensity associated with manually reading, coding and synthesizing large volumes of text (8). Analysts may require days or weeks to distill insights from submissions, particularly when technical depth and sector-specific terminology complicate interpretation. This constraint delays policymaking and reduces the ability of institutions to provide timely responses or iterations.

A second challenge relates to the lack of consistency across human reviewers. Manual summaries are shaped by the background, cognitive bias and interpretive style of the analyst. As a result, two reviewers evaluating the same set of submissions may produce summaries that differ considerably (9), a problem observed in previous consultation evaluations and reiterated in the project presentation slides. Such inconsistency threatens both the reliability and the perceived legitimacy of consultation outcomes.

The third limitation involves cognitive overload. Consultation feedback often contains mutually conflicting arguments, implicit assumptions, emotional expressions, or minority viewpoints that are easily overshadowed by more frequent contributions (10). Analysts may experience difficulty in maintaining a holistic view of these diverse perspectives, leading to unintentional underrepresentation of certain stakeholder groups. This issue becomes more pronounced as participation grows, reflecting a broader scalability problem.

Finally, in the context of the ELV Directive, the content of the submissions is deeply intertwined with complex technical domains, including battery chemistry, traceability systems, vehicle dismantling procedures and hazardous material regulations (11,12). Without advanced analytical tools, accurately synthesizing such multidimensional content is a formidable challenge. This serves as a compelling motivation to investigate whether generative AI can support the consultation process without compromising ethical standards or democratic inclusiveness.

### 3. Methodology

This study employed a multi-stage methodology combining data collection, preprocessing, LLM-based summarization and human evaluation. While the present study conducted these steps manually, the project documentation emphasizes that much of this pipeline, including comment extraction and data formatting, can be automated using Python-based tools such as BeautifulSoup, Pandas and regular expression filters.

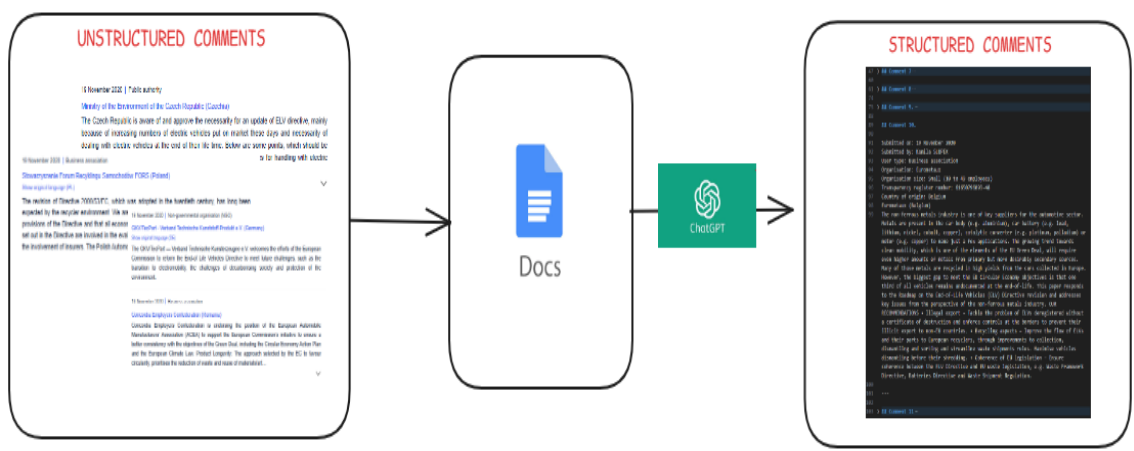

**Fig. 1:** LLM for Data Formatting Process

### 3.1 Data Collection and Preprocessing

Public submissions on the ELV proposal were extracted from the European Commission's online consultation portal. The corpus included contributions from automotive manufacturers, recycling firms, environmental NGOs, national authorities and individual citizens, resulting in a highly heterogeneous dataset. Text cleaning procedures were applied to remove redundant content, correct encoding errors and standardize formatting. Although multilingual submissions were included, the analysis focused on English-language texts or machine-translated equivalents to ensure uniformity.

### 3.2 LLM Prompting and Execution

Four large language models; Gemini, DeepSeek, Grok and Perplexity, were selected for analysis based on their performance characteristics and architectural diversity. Each model received standardized prompts instructing it to generate thematic summaries capturing major issues, distinguishing between stakeholder groups, identifying policy gaps and highlighting emergent concerns. Prompts were refined iteratively to emphasize neutrality, completeness and human-centered framing, consistent with the digital humanism principles attached to the project

The models also produced extended thematic analyses that mapped stakeholder arguments across clusters such as traceability, extended producer responsibility (EPR), dismantling procedures, recycled content mandates and material-specific recycling challenges.

### 3.3 Evaluation Framework

Human evaluators independently assessed each model's output along five dimensions: coverage, accuracy, balance, clarity and actionability. Scores were assigned on a scale of 1 to 5. Inter-rater reliability was calculated using Cohen's Kappa to ensure consistency. Ethical considerations, such as compliance with GDPR and the need for interpretability were incorporated throughout the analytical process.

### 4. Results

The LLM-generated summaries showed a remarkable level of alignment with the original submissions, accurately capturing the diversity of perspectives represented in the consultation. A central finding across all models was the broad support for modernizing the ELV Directive to reflect the rise of electric vehicles and the environmental challenges associated with high-voltage battery management. Stakeholders repeatedly emphasized the importance of harmonized dismantling standards and the necessity of robust battery traceability systems.

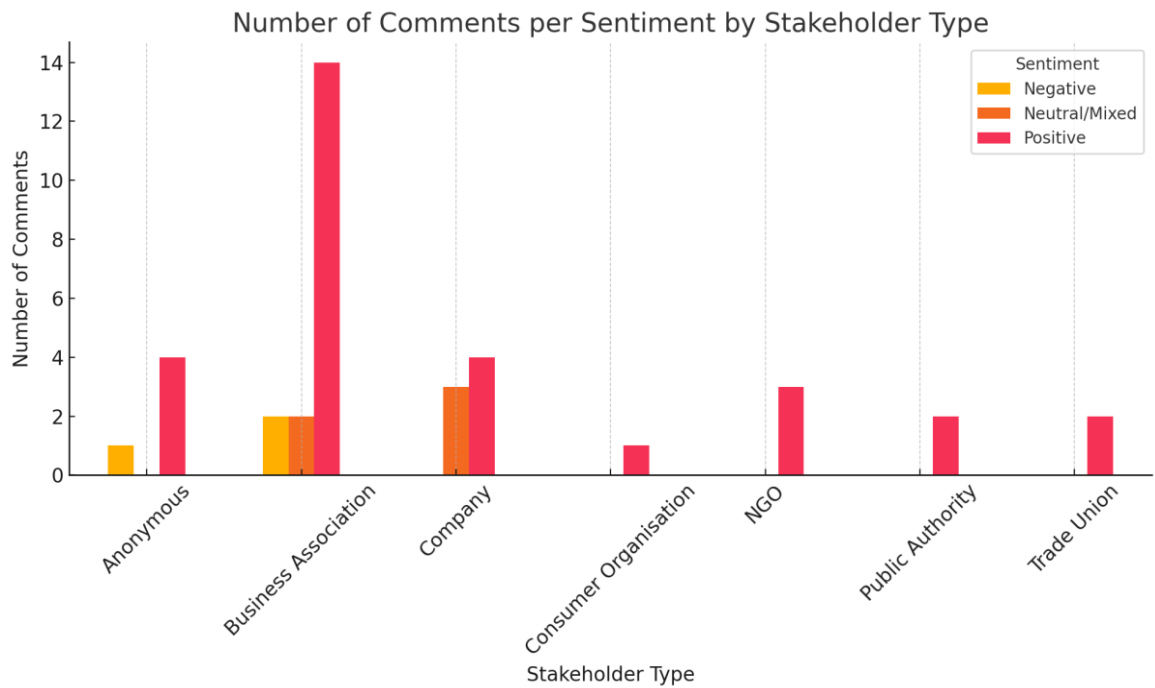

**Fig 2:** AI-Generated Analysis.

Stakeholders also expressed strong support for EU-wide traceability systems to address the longstanding issue of "missing vehicles" as seen in Fig. 2. LLM summaries highlighted stakeholder references to the Czech Republic's insurance-linked deregistration model and the Netherlands 'integrated vehicle registry as successful examples of traceability reforms. The models effectively differentiated between stakeholder groups, identifying divergent opinions on the implementation of EPR schemes. While NGOs advocated for centralized, government-supervised systems to ensure uniformity and accountability, industry actors favored more flexible, manufacturer-led mechanisms.

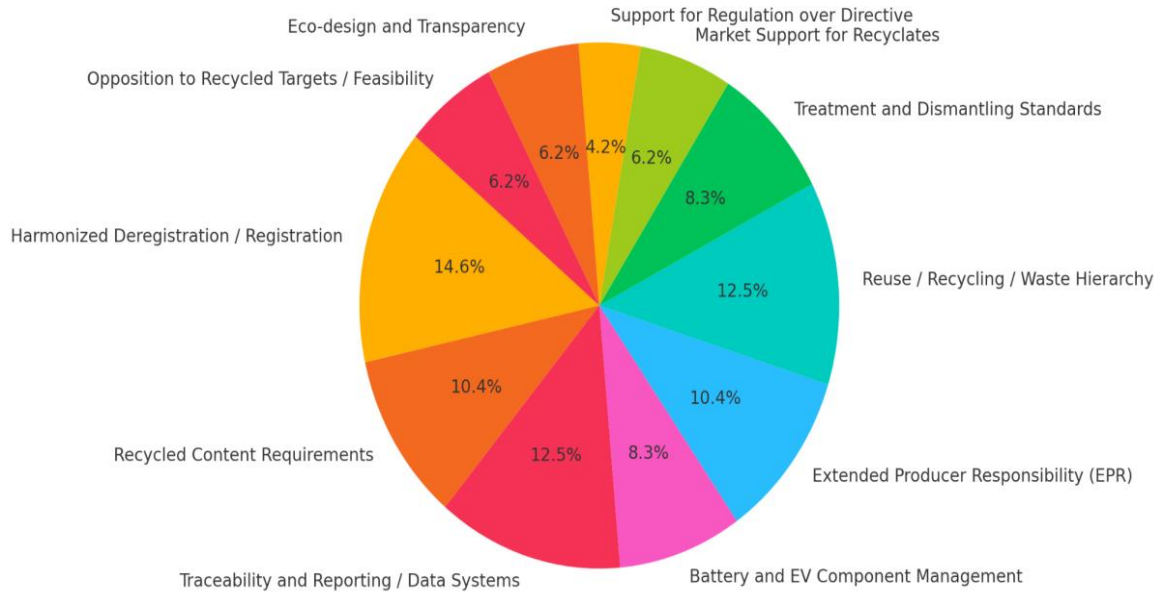

**Fig. 3:** Major themes analysis

Another salient theme was the debate around mandatory recycled content. LLMs accurately reflected the tension between industry concerns over supply chain limitations and the enthusiasm of environmental groups for more ambitious targets to stimulate the circular economy. Additionally, the models captured stakeholder demands for improved alignment between the ELV Directive and other EU regulatory frameworks, including the Batteries Regulation and REACH, an issue prominently highlighted in the project's thematic analysis slides

Quantitatively as seen in **Table 1** below; Perplexity consistently achieved the highest evaluation scores, particularly in balance and accuracy, reflecting its ability to incorporate factual grounding into summaries. Grok, while strong in thematic clustering, occasionally overlooked minority perspectives, resulting in lower actionability scores. Across all models, clarity and conciseness were consistently rated highly, indicating that the technology excels at producing readable and coherent summaries even when handling technically dense material.

| Metric | Coverage | Accuracy | Balance | Conciseness /Clarity | Actionability | Average |
|--------|----------|----------|---------|----------------------|---------------|---------|
| Gemini | 4 | 5 | 5 | 5 | 5 | **4.8** |
| DeepSeek | 4 | 5 | 4 | 5 | 5 | **4.6** |
| Grok | 4 | 4 | 5 | 5 | 4 | **4.4** |
| Perplexity | 5 | 5 | 5 | 5 | 5 | **5.0** |
| | | | | | | **4.7** |

**Table 1:** Evaluation of AI-Generated Summaries

**5. Discussion**

The findings from this study demonstrate that LLMs can significantly enhance the efficiency and inclusiveness of public consultation analysis. Generative AI reduces the cognitive and temporal burden associated with manually synthesizing submissions, enabling near real-time analysis, a benefit emphasized in the project briefing materials

This technological capability opens the possibility of dynamic consultation environments where policymakers can iteratively engage with citizens and refine proposals based on ongoing feedback.From an epistemic perspective, the models showed strong potential for identifying cross-cutting themes and surfacing both majority and minority viewpoints. Their performance suggests that LLMs can act as an analytical scaffold, supporting human experts in navigating complex datasets and avoiding interpretive blind spots.

The study reinforces the importance of hybrid governance models in which AI augments human decision-making rather than replacing it. This aligns closely with the Vienna Manifesto on Digital Humanism, which calls for technologies that support fairness, equity and human dignity. As the project slides emphasize, LLM-assisted consultation analysis can contribute to SDGs 9, 11 and 16 by promoting innovation, sustainable communities and strong institutions, a vision fully consistent with the long-term objectives of participatory environmental governance.

**6. Conclusion**

This research demonstrates that generative AI can play a meaningful role in improving the analysis of public consultation data. By producing high-quality summaries with strong thematic fidelity, LLLLMs offer a scalable and efficient method for distilling complex stakeholder input, particularly in technically demanding contexts such as the ELV Directive revision. However, the integration of AI must be approached with caution. Human oversight remains indispensable for ensuring factual reliability, ethical integrity and democratic legitimacy.

The future of public consultation lies not in replacing human analysts with automated systems, but in forging a participatory AI ecosystem where citizens are aware of how their contributions are analyzed and where policymakers leverage AI to deepen, not diminish the quality of democratic deliberation. With appropriate safeguards, AI can amplify public discourse and strengthen the inclusiveness and transparency of environmental policymaking.

## 7. Limitations

Although this study demonstrates the promise of generative AI in public consultation analysis, several limitations temper these findings. LLM outputs, while often accurate, remain probabilistic and may introduce fabricated or distorted details, especially when interpreting nuanced or technical submissions. Such hallucinations can undermine the precision required in legislative contexts.

A further limitation concerns representativeness: models tend to privilege widely expressed or linguistically straightforward viewpoints, which can result in the underrepresentation of minority or highly technical perspectives.

This imbalance was visible in some outputs, where less common but important concerns received minimal emphasis. Additionally, multilingual submissions introduce subtleties that machine translation may not fully preserve, potentially influencing AI interpretation.

The study also relied on a limited number of general-purpose LLMs that were not fine-tuned for EU environmental law. Consequently, the models sometimes overlooked deeper regulatory interdependencies. These limitations suggest that AI-generated insights should be treated as complementary and exploratory, requiring human validation and future methodological refinement.

## 8. Ethical Considerations

The use of generative AI in public consultations raises important ethical considerations that must guide its responsible integration. Transparency is essential: citizens and policymakers alike should know when AI tools are used, how summaries are produced and what data informs them, as this clarity is crucial for maintaining trust in digital governance. Equally important is accountability; AI outputs must not replace human judgment and policymakers must remain responsible for validating interpretations and ensuring that all stakeholder views, especially those from marginalized groups are fairly represented.

Privacy considerations also arise, as consultation data may include personal or sensitive information that must be processed in accordance with GDPR and broader data protection norms. Beyond this, explainability is necessary to prevent AI from becoming an opaque intermediary in democratic processes.

Policymakers should be able to trace AI-generated conclusions back to original submissions thus ensuring that decisions remain understandable, contestable and accountable to the public.

## 9. Future Work

Future research should focus on building more robust, scalable and domain-sensitive AI systems to strengthen the analysis of public consultations. A key direction is the development of automated pipelines capable of performing real-time data extraction, cleaning, thematic clustering and summarization, enabling policymakers to interact dynamically with evolving citizen feedback.

Advancing domain-specific fine-tuning is also essential; training models on EU environmental law, automotive circularity and prior consultations would improve factual grounding and reduce hallucinations. Additional work is needed to integrate multimodal inputs, including diagrams, audio submissions and technical datasets to create richer and more holistic analyses.

Equally important is the design of bias-mitigation mechanisms that ensure minority or infrequent viewpoints receive appropriate visibility.

Future systems should support participatory engagement by offering citizens tailored feedback, policy simulations, or explanatory tools that help them understand the impact of their input. Such enhancements would deepen transparency, promote co-creation and strengthen democratic dialogue in line with digital humanism principles.

## References

1. Waddington H, Sonnenfeld A, Finetti J, Gaarder M, John D, Stevenson J. engagement in public services in low-and middle-income countries: A mixed-methods systematic review of participation, inclusion, transparency and accountability …. Wiley Online Library [Internet]. 2019 Jun 1 [cited 2025 Dec 9];15(1–2). Available from: https://onlinelibrary.wiley.com/doi/abs/10.1002/cl2.1025

2. Sciences AMIR of A, 2011 undefined. Innovations in democratic governance: how does citizen participation contribute to a better democracy? journals.sagepub.com [Internet]. 2011 Jun 9 [cited 2025 Dec 9];77(2):275–93. Available from: https://journals.sagepub.com/doi/abs/10.1177/0020852311399851

3. Moynihan D, review JSP administration, 2014 undefined. Policy feedback and the politics of administration. Wiley Online Library [Internet]. 2014 [cited 2025 Dec 9];74(3):320–32. Available from: https://onlinelibrary.wiley.com/doi/abs/10.1111/puar.12200

4. Review WWPA, 2004 undefined. Formal procedures, informal processes, accountability, and responsiveness in bureaucratic policy making: An institutional policy analysis. Wiley Online Library [Internet]. 2004 Jan [cited 2025 Dec 9];64(1):66–80. Available from: https://onlinelibrary.wiley.com/doi/abs/10.1111/j.1540-6210.2004.00347.x

5. Gerrard J, production MKJ of cleaner, 2007 undefined. Is European end-of-life vehicle legislation living up to expectations? Assessing the impact of the ELV Directive on 'green'innovation and vehicle recovery. Elsevier [Internet]. [cited 2025 Dec 9]; Available from: https://www.sciencedirect.com/science/article/pii/S0959652605002593

6. Waddell S. Societal learning and change: How governments, business and civil society are creating solutions to complex multi-stakeholder problems [Internet]. Societal Learning and Change: How Governments, Business and Civil Society are Creating Solutions to Complex Multi-Stakeholder Problems. Taylor and Francis; 2017 [cited 2025 Dec 9]. 1–164 p. Available from: https://www.taylorfrancis.com/books/mono/10.4324/9781351280761/societal-learning-change-steve-waddell

7. Ke W, Zheng Y, Li Y, Xu H, Nie D, … PWAT on, et al. Large Language Models in Document Intelligence: A Comprehensive Survey, Recent Advances, Challenges, and Future Trends. dl.acm.org [Internet]. 2026 Jan 31 [cited 2025 Dec 9];44(1):1–64. Available from: https://dl.acm.org/doi/abs/10.1145/3768156

8. Benoit K, Conway D, … BLAP, 2016 undefined. Crowd-sourced text analysis: Reproducible and agile production of political data. cambridge.org [Internet]. [cited 2025 Dec 9]; Available from: https://www.cambridge.org/core/journals/american-political-science-review/article/crowdsourced-text-analysis-reproducible-and-agile-production-of-political-data/EC674A9384A19CFA357BC2B525461AC3

9. Cook MB, Smallman HS. Human factors of the confirmation bias in intelligence analysis: Decision support from graphical evidence landscapes. journals.sagepub.com [Internet]. 2008 Oct [cited 2025 Dec 9];50(5):745–54. Available from: https://journals.sagepub.com/doi/abs/10.1518/001872008X354183

10. Vuorenheimo M. Reducing Cognitive Biases in Business Intelligence: A Framework for Objective Data Analysis. Anal Chem [Internet]. 2025 Jun 16 [cited 2025 Dec 9];92(12):59. Available from: https://aaltodoc.aalto.fi/items/f54f3de4-683b-4451-aced-4c0a5a3f1bcc

11. Kostenko G, Energy AZSR in, 2024 undefined. World experience of legislative regulation for lithium-ion electric vehicle batteries considering their second-life application in power sector. systemre.org [Internet]. [cited 2025 Dec 9]; Available from: https://systemre.org/index.php/journal/article/view/836

12. Gianvincenzi M, Marconi M, Mosconi E, Sustainability CF, 2024 undefined. Systematic review of battery life cycle management: a framework for European regulation compliance. mdpi.com [Internet]. [cited 2025 Dec 9]; Available from: https://www.mdpi.com/2071-1050/16/22/10026

# Conclusions

When we planned the first African Digital Humanism Summer School, our goal was to bring together inspiring speakers, critical thinkers, curious explorers, and pragmatic practitioners for discussing Digital Humanism (Werthner, 2025) from African perspectives, all from very different contexts, experiences, and cultural backgrounds. Our participants and speakers represented all that great diversity. To stimulate active engagement, we defined several use cases from which the students could choose their group topic for the practical assignment. At the end of each day, we found the students in groups deeply diving into their subjects with such an energy and passion. It was a joy to witness the progress made over the days, leading to an inspiring concluding session of the summer school in Kigali, Rwanda. While the students' papers compiled here should not be considered as the final outcome of a scientific study, they show very well the diversity of questions, challenges, possible adoptions, approaches and solutions, when we try to find answers to how we should develop or how we should use AI technologies in the context of digital humanism. As normal in lively discourses, we may disagree with some of the statements, however, there is a fruitful richness and complementarity in the papers, reflected in these key findings:

Genuine cultural intelligence in AI requires not only linguistic processing but also socio-emotional sensitivity, demanding a paradigm shift toward language models that interpret meaning through relational, historical and communal lenses.

The future of the use of AI in administrative processes, such as public consultation, lies not in replacing human analysts with automated systems, but in forging a participatory AI ecosystem where citizens are aware of how their contributions are analyzed and where policymakers leverage AI to deepen, not diminish the quality of democratic deliberation. With appropriate safeguards, AI tools may amplify public discourse and strengthen the inclusiveness and transparency of policymaking and its feedback loops.

We want to underline the importance of combining AI capabilities with human expertise to produce high-quality, context-sensitive translations. Specific challenges are found in less-resourced languages, underlining the need for continued development and adaptation of AI tools.

We need to keep a critical view on AI generated content, but also make use of the advantages provided by AI technologies: Human-AI fact-checking is most successful when each side does what it does best: AI should handle bulk data processing, while humans apply judgment, context and creativity.

Future research should develop new languages models and fine-tune existing ones to better understand and process code-switched data like often present in African internet sources. Instead of treating code-switching as noise, these models should be designed to recognize it as a legitimate and important linguistic phenomenon, thereby creating technologies that are more relevant and useful in the context of African languages Furthermore, the models should integrate audio data. Currently there is a disconnect between written social media content and the vibrant, multilingual reality of spoken African languages. A promising avenue for data collection is to look into audio-based social media platforms, employing strong Automatic Speech Recognition (ASR) technology tailored to African languages. This enables us to capture a more authentic, conversational language use, including the nuances of intonation, rhythm and natural code-switching that are often lost in written text. This would provide a more representative dataset for building truly conversational LLMs.

To measure true progress in the quality of AI tools, we need to create benchmarks that are culturally and linguistically relevant. These benchmarks should not only test a model's ability to handle monolingual text but also its competence in managing tasks that involve code-switching, transliteration and context-specific cultural references. By developing and using these benchmarks, we can ensure that language technologies are not just performant, but also deeply respectful and useful for the communities they are intended to serve.

Our works highlight the need for AI systems that are technically capable and human-centered, inclusive and culturally aware. Ensuring equitable access requires expanding training data for major African languages, embedding cultural sensitivity into AI design and maintaining human oversight in practical applications. Building on our evaluation and the principles of Digital Humanism, next steps should focus on collaborating with local linguistic and cultural experts, advocating for government support to integrate African languages into AI and NLP tools, maintaining human oversight of AI outputs and promoting ethical, inclusive and culturally aware AI development.

Plenty of interesting challenges are ahead of us. Let us tackle them together, in cultural diversity united!

**Reference**

Hannes Werthner, Digital Humanism: On Digitalization and Artificial Intelligence (2025). Springer Cham, https://doi.org/10.1007/978-3-031-86905-1